\documentstyle[epsfig,aps,prd]{revtex} 
\def\laq{\raise 0.4ex\hbox{$<$}\kern -0.8em\lower 0.62 ex\hbox{$\sim$}}
\def\gaq{\raise 0.4ex\hbox{$>$}\kern -0.7em\lower 0.62 ex\hbox{$\sim$}}

\begin{document}
\bibliographystyle{unsrt}

\title{Production and detection of relic gravitons 
in quintessential inflationary models}

\author{Massimo Giovannini\footnote{Electronic address: 
giovan@cosmos2.phy.tufts.edu }}

\address{{\it Institute of Cosmology, Department of Physics and Astronomy,}}
\address{{\it Tufts University, Medford, Massachusetts 02155, USA}}

\maketitle

\begin{abstract}

A large class of quintessential inflationary models, recently 
proposed  by Peebles and Vilenkin, leads to 
post-inflationary phases whose effective equation of state 
is  stiffer than radiation. The expected gravitational 
waves logarithmic energy spectra are tilted towards high frequencies and 
characterized by two parameters: the inflationary curvature 
scale at which the transition to the stiff phase occurs and the 
number of (non conformally coupled) scalar degrees of freedom
 whose decay into fermions triggers the onset of a gravitational 
reheating of the Universe.
Depending 
upon the parameters of the model and upon the different inflationary 
dynamics (prior to the onset of the stiff evolution) the relic
gravitons energy density can be much more sizeable than in standard  
inflationary models, for frequencies larger than $1$ Hz.
We estimate the required sensitivity for detection of the 
predicted spectral amplitude and show that the allowed
 region of our parameter space 
leads to a signal smaller (by one $1.5$ orders of magnitude) 
than the advanced LIGO sensitivity 
at a frequency of $0.1$ KHz. The maximal signal, in our context,
is expected in the GHz region where the energy density of relic 
gravitons in critical units (i.e. $h_0^2 \Omega_{{\rm GW}}$) 
is of the order of $10^{-6}$, roughly eight orders 
of magnitude larger than in ordinary inflationary models.
Smaller detectors (not necessarily interferometers) 
can be relevant for detection purposes in the GHz frequency window. 
We suggest/speculate that future measurements through microwave cavities  
 can offer interesting perspectives.  

\end{abstract}
\vskip0.5pc

\noindent

\renewcommand{\theequation}{1.\arabic{equation}}
\setcounter{equation}{0}
\section{Formulation of the Problem} 

The idea that our present Universe could be populated by a sea of 
stochastically distributed gravitational waves (GW) is both 
{\em experimentally appealing} and {\em theoretically plausible}.
It is {\em appealing} since it would offer a natural cosmological 
source for the GW detectors which will 
come in operation during the next 
decade, like LIGO \cite{ligo}, VIRGO \cite{virgo}, LISA \cite{lisa} and 
GEO-600 \cite{geo} \footnote{LIGO (Laser Interferometric Gravitational 
wave Observatory), LISA (Laser Interferometer Space Antenna)}.
It is also  {\em plausible}, since nearly all the models trying to 
describe the 
first moments of the life of the Universe do predict the formation of 
stochastic gravitational wave backgrounds \cite{rev,vil}.

Our knowledge of early the Universe is only 
indirect. The success of big-bang 
nucleosynthesis
 (BBN) offers an explanation of the existence
of light elements whose abundances 
are of the the same order in different and distant galaxies. 
BBN hints that when the cosmic  plasma  was as hot as 
$0.1$ MeV, the Universe was probably dominated by radiation \cite{pee}. 
Prior to this moment direct cosmological observation are lacking but 
one can be reasonably confident that the laws of physics 
probed in particle accelerators still hold.  
Almost ten years of LEP (Large Electron Postitron collider) tested 
the minimal standard model (MSM) of particle interactions 
to the precision of the one per thousand for center of mass energies 
of the order of the $Z$ boson resonance. The cosmological implications 
of the validity of the MSM are quite important especially
 for what concerns the problem of the baryon asymmetry 
of the Universe and of the electroweak 
phase transition \cite{mis}. In spite of the success of the 
MSM we have neither direct nor indirect hints concerning the evolution 
of the Universe for temperatures higher than $100$ GeV. 
The causality principle applied  to the Cosmic Microwave 
(CMB) phtons seems to demand a moment where 
different patches of the Universe emitting a highly isotropic CMB 
were brought in causal contact. This is one of the original motivations  
of the inflationary paradigm \cite{infl}.

It is not unreasonable to think that in its early stages the 
Universe passed through different rates of expansion deviating 
(more or less dramatically) from the radiation dominated evolution.
It has been correctly pointed out through the years  and in different
frameworks \cite{gris1}
that every change in the early history of the Hubble parameter leads, 
inevitably, to the formation of a stochastic gravitational wave 
spectrum whose frequency behavior can be used in order to reconstruct 
the thermodinamical history of the early Universe.
The question which  naturally arises concerns the  strength
of  the produced gravitational wave background.

\begin{figure}
\centerline{\epsfxsize = 8 cm  \epsffile{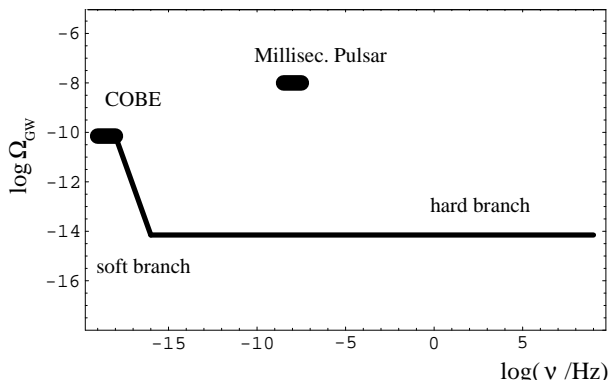}} 
\caption[a]{We report the graviton spectrum computed in the case of a 
pure de Sitter phase evolving towards  the (present) matter dominated 
phase through an intermediate radiation dominated 
stage of expansion. 
The spectrum has a soft branch ( $10^{-18} ~
{\rm Hz} < \nu < 10^{-16}~ {\rm Hz}$ ) and a hard branch ( $10^{-16}~ 
{\rm Hz} < \nu < 10^{9} ~{\rm Hz}$). In the 
two black boxes we spot 
 the COBE and the  millisecond pulsar timing bound.
Only for illustration purposes we plotted the spectrum for the 
largest amplitude consistent with Eq. (\ref{cobe}).}
\label{f1}
\end{figure}

If an inflationary phase 
is suddenly followed by a radiation dominated phase 
preceding the  matter dominated epoch, 
the amplitude of the 
 produced gravitons background can be computed  
and the result is illustrated  in Fig. \ref{f1},  where
we report the logarithmic energy spectrum of relic gravitons 
\begin{equation}
\Omega(\nu, \eta_0) = \frac{1}{\rho_{{\rm c}}} 
\frac{d \rho_{{\rm GW}}}{d \ln \nu},
\label{def}
\end{equation}
at the present (conformal) time $\eta_0$  as a function of the 
frequency $\nu$ ( $\rho_{{\rm GW}}$ is the
energy density of the produced gravitons and $\rho_c$ 
is the critical energy
density) \footnote{ Notice that in this paper we will denote 
with $\ln{}$ the Neperian logarithm and with $\log{}$ the logarithm in ten 
basis.}.
 
Since the energy spectrum ranges over several orders of magnitude 
it is useful to plot energy density per logarithmic interval of frequency. 
The spectrum consists of two branches 
a {\em soft} branch ranging  between 
$\nu_0 =1.1 \times~ 10^{-18}~h_0$ Hz
(corresponding to the present horizon) and 
$\nu_{{\rm dec}} = 
1.65 \times 10^{-16}~(\Omega_0~h_0^2)^{1/2}$ Hz 
(where $\Omega_0$ is the present fraction of critical density 
in matter and 
$0.5<h_0<1$ is the indetermination 
in the experimental value of the Hubble constant). For 
$\nu >\nu_{{\rm dec}}$ we have instead the {\em hard}
 branch consisting of high frequency gravitons mainly produced 
thanks to the transition from the inflationary regime to radiation. In the 
soft branch $\Omega_{{\rm GW}}(\nu,\eta_0) \sim \nu^{-2}$. 
In the hard branch $\Omega_{{\rm GW}}(\nu,\eta_0) $
 is  constant in frequency (or almost constant in the quasi-de Sitter case 
[see Section VI]). The soft branch was computed for the first 
time in \cite{rub} (see also \cite{ab}). The hard branch has been 
computed originally in \cite{star} (see also \cite{inflsp}).  
The COBE (Cosmic Microwave Background Explorer) observations of the first 
(thirty) multipole moments of the temperature fluctuations in the 
microwave 
sky imply \cite{cobe}
 that the gravitational wave contribution to the Sachs-Wolfe 
integral 
cannot be larger than the (measured) amount of anisotropy directly
detected. The soft branch 
of the  spectrum is then constrained and  the bound reads
\begin{equation}
\Omega_{{\rm GW}}(\nu,\eta_0)h^2_{0} \laq~ 6.9~ \times 10^{-11},
\label{cobe}
\end{equation}
for  $\nu\sim \nu_{0}$.
Moreover, the very small size of the fractional timing error in the 
arrivals of the millisecond plusar's pulses imply that also the hard
branch is bounded according to 
\begin{equation}
\Omega_{{\rm GW}}(\nu, \eta_0)~ \laq~ 10^{-8},
\label{puls}
\end{equation}
for $\nu\sim 10^{-8}$ Hz corresponding, roughly, to the inverse of the 
observation time during which the various millisecond pulsars have been 
monitored \cite{taylor}.

The two constraints of Eqs. (\ref{cobe}) and (\ref{puls}) are reported 
in Fig. \ref{f1}, at the two relevant frequencies, with black boxes.
In Fig. \ref{f1} we have chosen to normalize the logarithmic 
energy spectrum to the largest possible amplitude consistent with the COBE 
bound.
The COBE and millisecond pulsar constraints are {\em differential}
since they limit, locally, the logarithmic derivative of the 
gravitons energy density. There exists also an {\em integral}  
bound coming from standard BBN analysis \cite{schw,kt} 
and constraining the integrated graviton energy spectrum: 
\begin{equation}
h^2_{0}\int_{\nu_{{\rm n}}}^{\nu_{{\rm max}}} \Omega_{{\rm GW}}(\nu,\eta_0) d\ln{\nu}
\laq ~ 0.2 ~\times~ 10^{-5},
\label{NS}
\end{equation}
where $\nu_{{\rm max}}$ 
corresponds to the (model dependent) ultra-violet cut-off 
of the spectrum and $\nu_{{\rm n}}$ is the frequency corresponding 
to the horizon scale at nucleosynthesis
 \footnote{Notice 
that the BBN 
constraint of Eq. (\ref{NS}) has been derived in the 
context of the simplest BBN model, namely, 
assuming that no inhomogeneities and/or matter anti--matter domains 
are present at the onset of nucleosynthesis. In the presence of  
matter--antimatter domains for scales comparable with the 
neutron diffusion scale \cite{mm,jr} this bound might be slightly 
relaxed.}. It should be noted, in fact, that modes re-entering 
after the completion of nucleosynthesis will not increase 
the rate of the Universe expansion at earlier epochs.
From Fig. \ref{f1} we see that also  the global bound of 
 Eq. (\ref{NS})  is satisfied and the typical amplitude of the 
logarithmic energy spectrum in critical units 
for frequencies $\nu_{I} \sim 100 $ Hz (and larger)
cannot exceed $10^{-14}$. This amplitude 
has to be compared with the LIGO sensitivity to a flat
$\Omega_{{\rm GW}}(\nu_{I}, \eta_0)$ 
which could be {\em at most} of the order of 
$h_{0}^2 \Omega_{{\rm GW}}(\nu_{I},\eta_0)
= 5\times 10^{-11}$ after four months of
 observation with $90\% $ confidence (see third reference in 
\cite{rev}).
\begin{figure}
\centerline{\epsfxsize = 8 cm  \epsffile{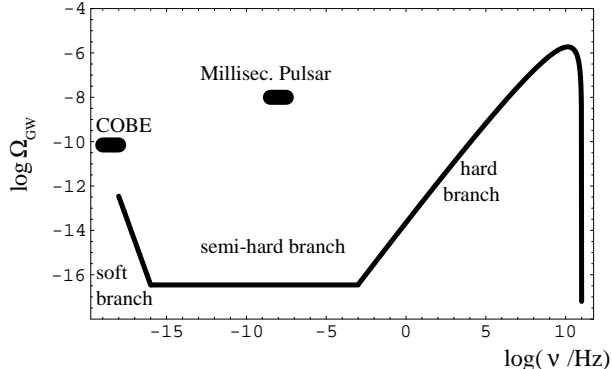}} 
\caption[a]{We illustrate the graviton spectrum produced 
by a pure de Sitter phase evolving towards a stiff phase. The hard 
branch corresponds to modes leaving the horizon during the 
de Sitter epoch and re-entering during the stiff phase. The soft and the 
 semi-hard branch are  made of gravitons re-entered, respectively,
 during the matter and during the radiation dominated phase. As in the
 previous figure we report  the various constraints on the 
differential spectrum. The frequency at which the spike starts developing 
is model dependent.}
\label{f2}
\end{figure}
Suppose that the hard branch
of the spectrum, reported in Fig. \ref{f1}, can be split into two further 
branches, a truly {\em hard branch} with growing slope 
and an intermediate {\em semi-hard} branch. 
The situation we are describing is indeed reproduced in Fig. 
\ref{f2} where 
the semi-hard branch now corresponds to the flat plateau and the hard 
branch to the spike associated with a broader peak. 
This class of spectra can be obtained  in 
the context of inflationary models 
{\em provided} the inflationary phase is followed by a 
a phase whose effective equation of state is stiffer than radiation. 
A model of this type has been recently investigated in Ref. \cite{pv} by 
J. Peebles and A. Vilenkin.

If an inflationary phase is
 followed by a stiff phase then, as it was showed in \cite{mm2}, one 
can indeed get a three branch spectrum including the usual 
soft and flat branches but supplemented by a truly hard spike. 
In general the slope of the logarithmic energy spectrum 
is typically ``blue''
since it mildly increases with the  frequency. 
More specifically the  slope depends upon the 
 stiff model and it can be shown \cite{mm2} that the maximal slope
 (corresponding to a linear increase in $\Omega_{{\rm GW}}(\nu,\eta_0)$) 
can be achieved in the 
case where the sound velocity of the effective matter sources 
exactly equals the speed of light \cite{zel1,zel2}.
In Fig. \ref{f2} 
we illustrate the 
case of maximal slope in the hard branch corresponding to 
$\Omega_{{\rm GW}}(\nu,\eta_0) \sim \nu \ln{\nu}$. 

Given the flatness of the spectra arising in the case 
of ordinary inflationary models (see Fig. \ref{f1}) the 
most constraining bound comes from large scale observations.
In our case the most constraining bounds for the height 
of the spike and for the whole spectrum come from short
distance physics and, in particular, from Eq. (\ref{NS}). 
In order to visually motivate the need for an accurate computation
of the graviton spectra in the case where an inflationary phase 
is followed by a stiff phase,   
let us  focus our attention on the 
frequency range where the gravitational wave detectors are (or  will be)
operating.
\begin{figure}
\centerline{\epsfxsize = 8 cm  \epsffile{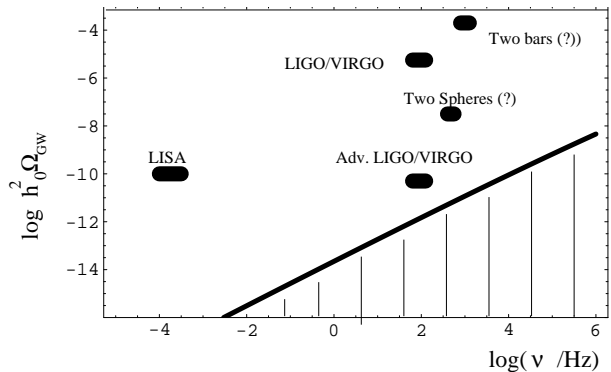}} 
\caption[a]{ We draw the hard branch of the relic graviton energy density
reported in 
Fig. \ref{f2} with particular attention to frequency range where the 
planned gravitational wave detectors are (or will be) operating. 
The dashed region marks the portion of the 
spectrum of Fig. \ref{f2} which is {\em above} the one of Fig. 
\ref{f1}. }
\label{f3}
\end{figure}
From Fig. \ref{f3} we see that around the LIGO 
\cite{ligo} and VIRGO \cite{virgo} 
frequency the hard branch of the spectrum has a larger amplitude 
if compared to the case of the spectral amplitude obtained when a pure 
de Sitter phase evolves suddenly towards a radiation dominated epoch.

In Fig. \ref{f3} we also illustrate (with thick black) boxes the 
expected sensitivities for interferometric detectors 
(LIGO/VIRGO) and for their advanced versions.  
In the same figure we also report the expected sensitivities coming from the 
cryogenic, resonant-mass detector  EXPLORER, while operating in CERN at a 
frequency of $923$ Hz \cite{bar}. EXPLORER provided a bound on 
$\Omega_{GW} \laq 500$ which is clearly to high  to  be of cosmological 
interest. Nevertheless, by cross-correlating  the data obtained from 
bar detectors (EXPLORER, NAUTILUS, AURIGA) it is not 
unreasonable to expect a sensitivity as large as $\Omega_{GW} \sim 10^{-4}$
in the KHz region.  Notice finally  that the cross-correlation of 
two resonant spherical detectors \cite{sph} might be able  to achieve 
sensitivities as low as $\Omega_{GW}\sim 10^{-7} $ always in the KHz 
range.

Spikes in the stochastic graviton background are not
forbidden by observations and are also theoretically plausible 
whenever a stiff phase follows a radiation dominated phase. 
In this paper, by complementing and extending the analysis 
of \cite{pv} and of \cite{mm2}, we want to study 
more accurately the spectral 
properties of the relic gravitons with special 
attention to the structure of the hard peak. We will 
also be interested in comparing the predictions 
of the models with the foreseen capabilities 
of the interferometric and resonant detectors.

The plan of our paper is  the following. In Section II we will introduce 
the basic aspects of the quintessential inflationary models. In Section 
III we will compute the relic graviton energy spectra. 
In Section IV we will discuss the power spectra and the 
associated spectral densities.   
In Section V we will compare the obtained spectra with the sensitivities
of the planned interferometric and non-interferometric detectors. 
In Section VI we will analyze the impact 
of the  slow-rolling corrections on the structure of the hard peak.
Section VII contains our concluding remarks. For sake of 
completeness we made the choice of 
reporting in the Appendix some relevant derivations of the formulas 
used in obtaining our results.

\renewcommand{\theequation}{2.\arabic{equation}}
\setcounter{equation}{0}
\section{Quintessential Inflationary Models} 

Recently J. Peebles and A. Vilenkin 
\cite{pv}   presented a model where the idea 
of a post-inflationary phase stiffer 
than radiation is dynamically realized.
One of the motivations of the scenario
 is related to a recent set of observations which seem to suggest 
that $\Omega_{0}$ (the present density parameter in baryonic plus  
dark matter) should be significantly smaller than one and 
probably of the order of $0.3$. If the Universe is flat, the 
relation between luminosity and red-shift observed for Type Ia supernovae
\cite{obs}
seem to suggest that the missing energy should be stored in a fluid 
with negative pressure. The missing energy stored in this fluid should be 
of the order of $10^{-47}$ GeV$^4$, too small if compared with the
cosmological constant arising from electroweak spontaneous symmetry bereaking
(which would contribute with $(250 ~{{\rm GeV}})^4$). The idea is that 
this effective cosmological constant could come from 
a scalar field $\phi$ (the quintessence \cite{quint} field) 
whose potential is unbounded from below \cite{rp}. 
According to Peebles and Vilenkin, 
$\phi$ could be identified with the inflaton and, as 
a result of this identification, the effective potential of 
$\phi$  will inflate for $\phi< 0$ and it will be unbounded 
from below for $\phi \geq 0$ acting, today, as an effective (time dependent)
comological term. A possible potential leading to the mentionad dynamics 
could be 
\begin{eqnarray}
V(\phi) = \lambda ( \phi^4 + M^4) ,~~{\rm for}~\phi<0,~~~{\rm and}~~~
V(\phi) = \frac{\lambda M^8}{\phi^4 + M^4},~~{\rm for}~~\phi\geq 0.
\label{pot}
\end{eqnarray}
where, if we want the present energy density in $\phi$ 
to be comparable with (but less then)  the total 
(present) energy density, we have to require $M\sim 10^{6}$ GeV. 
The scenario we are describing can be implemented with any other 
inflationary potential (for $\phi<0$) and 
 the example of a chaotic potential is 
only illustrative. 
Our considerations will be largely independent on the 
specific potential used and we will comment, when needed, 
aboout possible differences  induced by the specific 
type of potential. 

Let us consider the evolution equations of an inflationary Universe 
driven by a single field $\phi$ in a conformally flat 
metric 
\begin{equation}
ds^2 = a^2(\eta)( d\eta^2 - d\vec{x}^2).
\label{metric}
\end{equation}
Using the  conformal time $\eta$ 
the coupled system describing the evolution of the scale
factor and of $\phi$ is
\begin{eqnarray}
&&M_{P}^2 {\cal H}^2 = \biggl(\frac{\phi'^2}{2} + a^2 V(\phi)\biggr),
\nonumber\\
&& M_{P}^2 {\cal H}' = -\biggl( \phi'^2 -  a^2 V(\phi) \biggr),
\nonumber\\
&& \phi'' + 2 {\cal H} \phi' + a^2 \frac{ \partial V}{\partial\phi}=0,
\label{eqm}
\end{eqnarray}
where $l_{P} = M_{P}^{-1} = \sqrt{8\pi G/3}$.
Since the scalar field potential is unbounded from below, 
after a phase of slow-rolling the inflaton 
evolves towards a phase where the kinetic energy of the inflaton 
dominates. For instance one can bear in mind the form of $V(\phi)$ 
reported in Eq. (\ref{pot}). The background enters then a 
stiff phase where the energy density of the inflaton $\rho_{\phi}$ 
and the scale factor evolve as 
\begin{equation}
\rho_{\phi} = H_1^2 M_{P}^2 \biggl(\frac{a_1}{a}\biggr)^6 , ~~~
a(\eta) \sim \sqrt{\frac{\eta}{\eta_1}},~~~{\rm where},~~~
H_1= \frac{1}{a_1\eta_1} \sim  \sqrt{\lambda} M_{P}.
\end{equation}
When 
an inflationary phase  is followed by a stiff phase 
a lot of hard gravitons will be generated. At the same time the 
energy density of the background sources will decay as $a^{-6}$
whereas the energy density of the short wave-lengths gravitons will 
decay as $a^{-4}$.
Te Universe will soon be dominated by hard 
gravitons whose non-thermal 
spectrum \cite{mm2} would 
be unacceptable since gravitons cannot thermalize 
below the Planck scale. A solution to this potential difficulty 
 came from L. Ford \cite{ford} who noted that in the limit of nearly 
conformal coupling also scalar degrees of freedom 
(possibly coupled to fermions) are amplified.
If $N_{s}$ minimally coupled scalar field are present they can reheat 
the Universe with a thermal distribution since their energy spectra, 
amplified because of the transition from the inflationary to the stiff phase,
can thermalize thanks to  non-graviational (i.e. gauge) interactions which get 
to local thermal equilibrium well below the Planck energy scale.
It can be also shown that the same discussion can be carried on in the case 
where the scalar degrees of freedom are simply non conformally 
coupled \cite{dv}.

Suppose indeed that during the inflationary phase 
various scalar, tensor and vector degrees of freedom were present. 
Unless one adopts some rather contrived points of view we have 
to accept that, in Einsteinian theories of gravity, 
the only massless degrees 
of freedom to be amplified by a direct coupling 
to the background geometry are tensor fluctuations of the 
metric and 
non conformally coupled scalar fields since the evolutions 
equations of chiral fermions, gravitinos \cite{gravitinos} 
and gauge fields \cite{mm3}
are invariant under a Weyl rescaling of the metric tensor in 
a conformally flat background geometry as the one 
specified in Eq. (\ref{metric}). Of course, if the theory 
is not of Einstein Hilbert type this statement might 
be different.

The evolution equation of a non conformally coupled scalar field 
in a conformally flat FRW background reads
\begin{equation}
\psi'' + 2 {\cal H} \psi' - 6 \xi \bigl[ {\cal H}' +{\cal H}^2\bigr] 
- \nabla^2 \psi =0.
\end{equation}
By defining the corresponding proper amplitude $ \chi = a\psi$ we get 
that the previous equation can be written, in Fourier space, as 
\begin{equation}
\chi_{k}'' + \bigl[k^2 - V(\eta)\bigr] \chi_{k} =0,~~~~
V(\eta) = (1 - 6 \xi ) \frac{a''}{a},
\label{chi}
\end{equation}
where we see that the case of exact conformal coupling is recovered 
for $\xi\rightarrow 1/6$ whereas the case of minimal 
coupling occurs for $\xi \rightarrow 0$. A lot of work has been done 
in the past in order to compute the energy density of the quanta of 
the field $\psi$, excited as a result of the background geometry evolution
in the early stages of the life of the Universe \cite{birrel}. One can 
try to do the calculation either exactly (but only for rather specific forms 
of the effective potential of the Schroedinger-like equation (\ref{chi})), 
or approximately by identifying $|\xi -1/6|$ as the small parameter in
 the perturbative expansion. In this limit one can show \cite{ford,birrel}
that the energy density of the created quanta can be expressed as 
\begin{equation}
\rho_{\psi}(\eta) =\frac{1}{2 \pi^2 a^4} \int_{0}^{\infty}|\beta(k)|^2~k^3~ 
d k,~~~{\rm with}~~~ \beta(k) = \frac{i}{2 k} \int_{-\infty}^{+\infty} 
e^{- 2 i k\eta} V(\eta) d\eta.
\end{equation}
We can notice that in most of the examples we are interested in 
$V(\eta) \rightarrow \eta^{-2}$ for $\eta\rightarrow \pm\infty$. 
As an example we can consider $V(\eta) = q/(\eta^2 + \eta_1^2)$. Then 
by evaluating $\beta(k)$ with contour integration in the complex plane 
we can estimate that the energy density will be 
\begin{equation}
\rho_{\psi} = \frac{q^2( 1 - 6\xi)^2}{128} \frac{{\cal H}_1}{a^4} = 
  \frac{q^2( 1 - 6\xi)^2}{128} H_{1}^4, 
\end{equation}
where ${\cal H}_1 = q/\eta_1$, 
 and $H_1 = {\cal H}_1/a$ is the Hubble parameter 
in cosmic time.
Suppose now that during the inflationary phase there are  $N_{s}$ 
(minimally coupled and massless) scalar degrees of freedom $\psi_i$. 
Because 
of the minimal coupling to the geometry these scalar degrees of freedom 
will clearly be excited since their evolution equation is {\em not} 
invariant under conformal rescaling of the metric tensor.
The produced quanta associated with each $\psi_i$ can be 
computed by specifying (for each field mode) 
the initial  vacuum state deep in the de Sitter epoch and by 
ensuring a sufficiently smooth transition between the de Sitter and the 
stiff phase. We will perform a similar calculation for the case of 
GW in the next Section. Here we only report the main 
result which was originally obtained in  \cite{ford} in the case 
of quasi-conformal coupling (i.e. $|\xi -1/6|< 1$) and subsequently 
generalized to the case of generic $\xi$ in \cite{dv}:
\begin{equation}
\rho_{m}(\eta)= \sum_{i=1}^{N_{s}} \rho_{\psi_{i}}(\eta) \sim
R H_{1}^4 \biggl(\frac{a_1}{a}\biggr)^4,~~~{\rm where}~~~
R\sim  R_{i} N_{s}.
\label{rm}
\end{equation}
$R_{i}$ is the contribution of each massless scalar degree of freedom
to the energy density of the amplified fluctuations and it is of the order 
of $10^{-2}$.
The appearance of $H_1$ in the final expression of the energy of the 
created quanta can be simply understood since the typical spectra 
obtained in the transition from a de Sitter phase to a stiff phase 
are increasing in frequency \cite{mm2} and, therefore, the maximal
contribution to the energy density will come from the ultra-violet 
branch of the spectrum.

The creation of massless quanta of the fields 
$\psi_i$ triggers an interesting possibility of gravitational reheating. 
Since $\rho_{m}$ red-shifts faster than $\rho_{\phi}$ we have that 
there  will be a moment, $\eta_{r}$
 where the two energy densities will be of the 
same order. In the context of the present model this moment defines 
the onset of the radiation dominated phase. We can 
compute this moment by requiring that $\rho_{m}(\eta_r) \sim 
\rho_{\phi}(\eta_{r})$. The result is that 
\begin{equation}
\biggl(\frac{a_1}{a_{r}}\biggr) \sim \sqrt{R} \sqrt{\frac{H_1}{M_{P}}}
\sim 10^{-7} \sqrt{R},
\end{equation}
where we used the fact that in order to be compatible with the COBE
observation $\sqrt{\lambda} \sim 10^{-7}$ \cite{ks}. 
In view 
of  our application to GW it is interesting to compute 
the typical (present)
frequency at which the transition to radiation occurs. 
By red-shifting the curvature scale at $\eta_{r}$ (i.e. 
$H_{r} \equiv H(\eta_{r}) = H_1 (a_1/a_r)^3$) from $\eta_{r}$ up to 
now we obtain
\begin{equation}
\nu_{r}(\eta_0) = 3.58 ~R^{\frac{3}{4}}~ \bigl(\frac{\lambda}{10^{-14}}\bigr)~
\bigl(\frac{g_{\rm dec}}{g_{{\rm th}}}\bigr)^{1/3} {\rm mHz}, 
\end{equation}
where $g_{{\rm th}}$ 
is the number of spin degrees of freedom contributing to the 
thermal entropy after matter thermalization.
Amusingly enough this frequency is of the same order of the typical 
frequency of operation of LISA.
It is also interesting to compute the present value of the frequency 
$\nu_1$, with the result that
\begin{equation}
\nu_{1}(\eta_0) = 358~ R^{-\frac{1}{4}} 
\bigl(\frac{g_{\rm dec}}{g_{{\rm th}}}\bigr)^{1/3}~{\rm GHz}.
\end{equation}
The thermalization of the created quanta of the fields $\psi_i$ occurs quite 
rapidly, and its specific time is fixed by the moment 
at which the interaction rate becomes comparable with the Hubble expansion 
rate during the stiff phase. The typical energy of
the created quanta is of the order of $\epsilon\sim H_1 (a_1/a)$. The 
particle density is of the order of $n\sim R\epsilon^3$. Assuming  
that the created quanta interact through the exchange of 
gauge bosons, then the typical interaction cross  section will be
of the order of $\sigma\sim \alpha^2/\epsilon^2$. Thus, imposing 
that at thermalization $n(\eta_{\star}) \sigma(\eta_{\star})
 \sim H(\eta_{\star})$ we get that $a_{\star} /a_1\sim \alpha^{-1} R^{-1/2}$,
 with $\alpha\sim 10^{-1}$--$10^{-2}$.
The typical temperature associated with the transition from the 
stiff to the radiation dominated phase can be computed and it turns out to be
\begin{equation}
T_{r}= \bigl(\frac{H_{1}}{M_{P}}\bigr) R^{3/4} M_{P}\simeq 10^{3}~N_{s}^{3/4}
~~{\rm GeV}.  
\end{equation}
If we do not fine-tune $H_1$ to be much smaller than $10^{-7}$ in Planck units 
and if we take into account that $N_{s}$ has to be typically large in order 
to be compatible with standard BBN (see also Section III) we have to conclude 
that $T_{r}$ is typically a bit larger than $1$ TeV. 

\renewcommand{\theequation}{3.\arabic{equation}}
\setcounter{equation}{0}
\section{Gravitons Energy Spectra} 

We can characterize a generic graviton background 
in terms of three related 
(and equally important) physical observables. 
We can  
compute the (present)  spectral 
energy density in critical units $\Omega_{{\rm GW}}(\nu,\eta_0)$, but,
for experimental applications, two other quantities can be defined, namely 
the {\em power  spectrum} ( which will be denoted with 
$\delta_{h}(\nu,\eta_0)$)
and the {\em spectral density} $S_{h}(\nu,\eta_0)$. 
$\Omega_{{\rm GW}}(\nu,\eta_0)$ 
and $\delta_{h}(\nu,\eta_0)$ 
are dimension-less whereas the spectral density is measured 
in seconds. 
In Appendix A we give the precise mathematical definitions 
of these observables.  

The continuity  of the scale factors and of their first derivatives 
implies that the evolution of our model can be expressed as 
\begin{eqnarray}
&&a_{i}(\eta) = \biggl[-\frac{\eta_1}{\eta}\biggr],~~~~~~~~~~~~~~~~~
~~~~~~~~~~~~~~~~~~~~~~~~~~~~~~~~~~~~~~\eta \leq -\eta_1,
\nonumber\\
&&a_{s}(\eta)= \sqrt{\frac{ 2\eta + 3 \eta_1}{\eta_1} },~~~~~~~~
~~~~~~~~~~~~~~~~~~~~~~~~~~~~~~~~~~~~~~~-\eta_{1}<\eta \leq \eta_r,
\nonumber\\
&&a_{r}(\eta) =
 \frac{\eta+ 3 \eta_1 + \eta_r}{\sqrt{\eta_1 ( 2 \eta_r + 3\eta_1)}},
~~~~~~~~~~~~~~~~~~~~~~~~~~~~~~~~~~~~~~~~~~~\eta_{{\rm dec}}\leq\eta<\eta_{r},
\nonumber\\
&&a_{m}(\eta) = \frac{(\eta + \eta_{{\rm dec}} + 6\eta_1 
+ 2 \eta_{r})^2}{2( 2 \eta_{{\rm dec}} + 2 \eta_{r} 
+ 6 \eta_1) \sqrt{\eta_{1}( 2 \eta_r + 3\eta_1)}},~~~~~~~~~~~~~~~~~~~
\eta_{0} \leq \eta<\eta_{{\rm dec}},
\label{scalefact}
\end{eqnarray}
where $\eta_0$ and $\eta_{{\rm dec}}$ are, respectively, the present time
 and the decoupling time, whereas $\eta_1$ and $\eta_{r}$ have  
been defined in the previous Section.

The graviton field operators can be decomposed as 
\begin{equation}
\hat{\mu}(\vec{x},\eta) = \frac{1}{(2\pi)^{\frac{3}{2}}} \int d^3 k \bigl[ 
\hat{\mu}(k,\eta) e^{i\vec{k}\cdot\vec{x}} + \hat{\mu}^{\dagger}(k,\eta) 
e^{- i\vec{k}\cdot\vec{x}}\bigr],
\end{equation}
where $\hat{\mu}(k,\eta)= \psi(k,\eta) \hat{a}(\vec{k})$. This decomposition 
holds for each polarization. In order to compute the energy density of the 
graviton background we have to solve the evolution of the mode function 
\begin{equation}
\psi''  + \bigl[ k^2 - \frac{a''}{a}\bigr]\psi =0,
\label{mode}
\end{equation}
in each of the four temporal regions defined by Eq. (\ref{scalefact}). 
Notice that $a''/a$ has a bell-like shape 
and it goes asymptotically as $\eta^{-2}$ 
in each phase of the background evolution. Thus $\psi$ will oscillate for 
$k\eta\gg 1$ but it will be parametrically amplified in the opposite 
limit (i.e. $k\eta < 1$). At
$k\eta\sim 1$ the given mode will hit the potential barrier represented by 
$|a''/a|$. 
The solution of Eq. (\ref{mode}) in the 
background of Eq. (\ref{scalefact}) is: 
\begin{eqnarray}
&&\psi_{i}(k,\eta) = \frac{p}{\sqrt{2 k}} \sqrt{x} H^{(2)}_{\nu}(x),
~~~~~~~~~~~~~~~~~~~~~~~~~~~~~~~~~~~~~~~~~~~~~~~~~~~~~~~~~~~\eta < -\eta_1,
\nonumber\\
&&\psi_{s}(k,\eta) = \frac{1}{\sqrt{2 k}} \sqrt{y} \bigl[ s^{\ast} A_{+}(k) 
H^{(2)}_{0}(y) + s A_{-}(k) H^{(1)}_{0}(y)\bigr],~~~
~~~~~~~~~~~~~~~-\eta_{1}<\eta <\eta_{r},
\nonumber\\
&&\psi_{r}(k,\eta) =\frac{1}{\sqrt{2 k}}\bigl[ 
B_{+}(k) e^{-i z} + B_{-}(k) e^{i z}\bigr],~~~~~~~~~~~~~~~~~~~~~~~~~~~~~~~~~~
~~~~~~\eta_{r}<\eta <\eta_{{\rm dec}},
\nonumber\\
&&\psi_{m}(k,\eta) = \frac{1}{\sqrt{2 k}} \sqrt{w} \bigl[ q^{\ast} c_{+}(k) 
H^{(2)}_{\mu}(w) + q c_{-}(k) H^{(1)}_{\mu}(w)\bigr],~~~~~~~~~~~~~~~~~~~
\eta_{{\rm dec}}<\eta <\eta_{0},
\label{solmode}
\end{eqnarray} 
where,
\begin{equation}
p= \sqrt{\frac{\pi}{2}} e^{- i \frac{\pi}{2}\nu} e^{-i\frac{\pi}{4}},
~~~s=\sqrt{\frac{\pi}{2}}e^{i\frac{\pi}{4}},~~~
q= \sqrt{\frac{\pi}{2}} e^{i \frac{\pi}{2}\mu} e^{i\frac{\pi}{4}},
\end{equation}
guarantee that the large argument limit of the Hankel functions 
$H^{(1,2)}_{\mu,\nu}$ is exactly the one required by the quantum mechanical 
normalization 
\footnote{Notice that we kept the Hankel 
indices  $\nu$ and $\mu$ generic. In the
 case of a pure de Sitter phase we would have $\nu = \frac{3}{2}$.}.
The arguments of $\psi$ are, respectively,
\begin{equation}
x= k\eta,~~~y= k(\eta + \frac{3}{2}\eta_1),~~~z= k \eta,~~~
w= k(\eta + \eta_{{\rm dec}} + 6 \eta_{1} + 2 \eta_{r}),
\end{equation}
and the  six mixing coefficients ($A_{\pm}(k)$, 
$B_{\pm}(k)$, $c_{\pm}(k)$) can be fixed by the six conditions 
obtained matching $\psi$ and $\psi'$ in $\eta= -\eta_1$, 
$\eta =\eta_{r}$ and $\eta=\eta_{{\rm dec}}$. The results 
of this  calculation are reported in Appendix C. For a generic amplification 
coefficient $\beta_{-}(\omega)$ the spectral energy density 
in the relic graviton
background is given by Eq. (\ref{endens}) 
\begin{equation}
\frac{d \rho_{GW}}{d\ln{\omega}} = \frac{\omega^4}{\pi^2} 
\overline{n}(\omega),~~~~
\overline{n}(\omega) = |\beta_{-}(\omega)|^2,~~~\omega=\frac{k}{a}= 2\pi \nu
\end{equation}
since, as it is well known and discussed in Appendix A, the square modulus of 
the mixing coefficient can be interpreted as the mean number 
of gravitons at a given frequency. Notice that $\omega$ is the physical 
wave-number. The relic graviton 
energy spectrum (in critical units) in each of the three branches 
is simply obtained by inserting $A_{-}(\omega)$ (i.e. Eq. (\ref{amin}), 
$B_{-}(\omega)$ (i.e. Eq. (\ref{bmin}) and $c_{-}(\omega)$ (i.e. Eq. (\ref{cmin})) 
into  Eq. (\ref{endens}).
The final result can be expressed as 
\begin{eqnarray}
&&\Omega_{GW}(\omega, \eta_0) = \Omega_{\gamma}(\eta_0)~ \varepsilon~\lambda ~
\biggl(\frac{\omega}{\omega_{r}}\biggr) ~
\ln^2{\biggl(\frac{\omega}{\omega_1}\biggr)},~~~~
~~~~~~~~~\omega_{r} < \omega < \omega_1,
\nonumber\\
&& \Omega_{GW}(\omega,\eta_0) =  \Omega_{\gamma}(\eta_{0})~
\frac{\pi}{4}~ \varepsilon~\lambda ~~
\ln^2{\biggl(\frac{\omega_r}{\omega_1}\biggr)},~~~~~~~~~~~~~\omega_{{\rm dec}}
 <\omega <\omega_{r},
\nonumber\\
&& \Omega_{GW}(\omega,\eta_0) = \Omega_{\gamma}(\eta_0)~ 
~\frac{\pi}{16}~ \varepsilon~\lambda
~\biggl(\frac{\omega_{{\rm dec}}}{\omega}\biggr)^2
~ \ln^2{\biggl(\frac{\omega_r}{\omega_1}\bigg)}
,~~~
\omega_0< \omega<\omega_{{\rm dec}},
\label{gravendens}
\end{eqnarray}
with
\begin{equation}
\nu_{{\rm dec}}(\eta_0)= 1.69\times ~10^{-16}
 \times (\Omega_{0}(\eta_0) h^2_{0})^{1/2}~{\rm Hz},~~~and~~~
 \nu_0(\eta_0) = 1.1\times 10^{-18} ~h_0~ {\rm Hz},
\end{equation}
where $\Omega_0(\eta_0)$ is the  fraction of critical density in matter.
Notice that 
\begin{eqnarray}
&&\varepsilon=2 R_{{\rm i}} 
\bigl(\frac{g_{{\rm dec}}}{g_{{\rm th}}}\bigr)^{1/3},~~~R_{{\rm i}}= 
\frac{81}{32 ~\pi^3} 
\nonumber\\
&&\Omega_{\gamma}(\eta_0) = 
\frac{\rho_{\gamma}(\eta_0)}{\rho_{{\rm c}}(\eta_0)} 
\equiv \frac{g_{0}\pi^2 }{30} 
\frac{T^4_{0}}{H^2_{0}M^2_{P}} =2.6 \times 10^{-5} ~h^{-2}_{0},
\end{eqnarray}
where $g_0=2$ and  $T_0= 2.73~{\rm K}$. $\Omega_{\gamma}(\eta_0)$
 is the fraction of critical energy density in the form 
of radiation at the present observation time. 
Notice that the dependence upon the number of relativistic degrees of freedom 
occurs since, unlike gravitons, matter thermalizes and then 
the ratio between the 
critical energy density and the energy density
stored in the relic graviton background is only approximately constant in the 
radiation dominated phase. 
 
The local (differential) bounds on the energy spectrum can be easily
satisfied. Indeed by taking $H_1/M_{P} \leq 10^{-7}$ the spectrum 
satisfies the COBE bound of Eq. (\ref{cobe}) and also the pulsar bound 
of Eq. (\ref{puls}). The {\em indirect} nucleosynthesis bound applies to the 
integrated spectrum and since in our case the spectral energy density 
increases sharply in the hard branch we have to conclude that the height of
 the peak cannot be too large. In order to prevent the Universe 
from expanding too fast at nucleosynthesis we have to demand
\begin{equation}
\int_{\nu_{{\rm n}}}^{\nu_{{\rm max}}} 
d\ln{\omega} \Omega_{{\rm GW}}(\omega,\eta_{{\rm n}}) <
\frac{7}{43}(N_{\nu} -3) 
\biggl[\frac{\rho_{\gamma}(t_{{\rm n}})}{\rho_{c}(\eta_{{\rm n}})}\biggr]
\end{equation}
Since the maximal number of massless
 neutrinos permitted in the context of the 
homogeneous and isotropic BBN scenario is bounded to be $N_{\nu} \leq 3.4$, 
we have that in our context the nucleosynthesis bound becomes 
\begin{equation}
\frac{3}{N_{s}} \biggl(\frac{g_{{\rm n}}}{g_{{\rm th}}}\biggr)^{1/3} < 0.07,
\label{req}
\end{equation}
where the factor of $3$ counts the two polarizations 
of the gravitons but also the quanta associated with the inflaton. 
The number of relativistic degrees of freedom after matter thermalization
is given, in the  MSM by $g_{{\rm th}}= 106.75$, whereas $g_{{\rm n}} = 10.75$.
Eq. (\ref{req}) implies that 
the number of (minimally coupled) scalar degrees of freedom 
will have to exceed $20$ as it can occur, for instance, in the 
minimal supersymmetric standard model \cite{pv}.

\begin{figure}
\begin{center}
\begin{tabular}{|c|c|}
      \hline
      \hbox{\epsfxsize = 7 cm  \epsffile{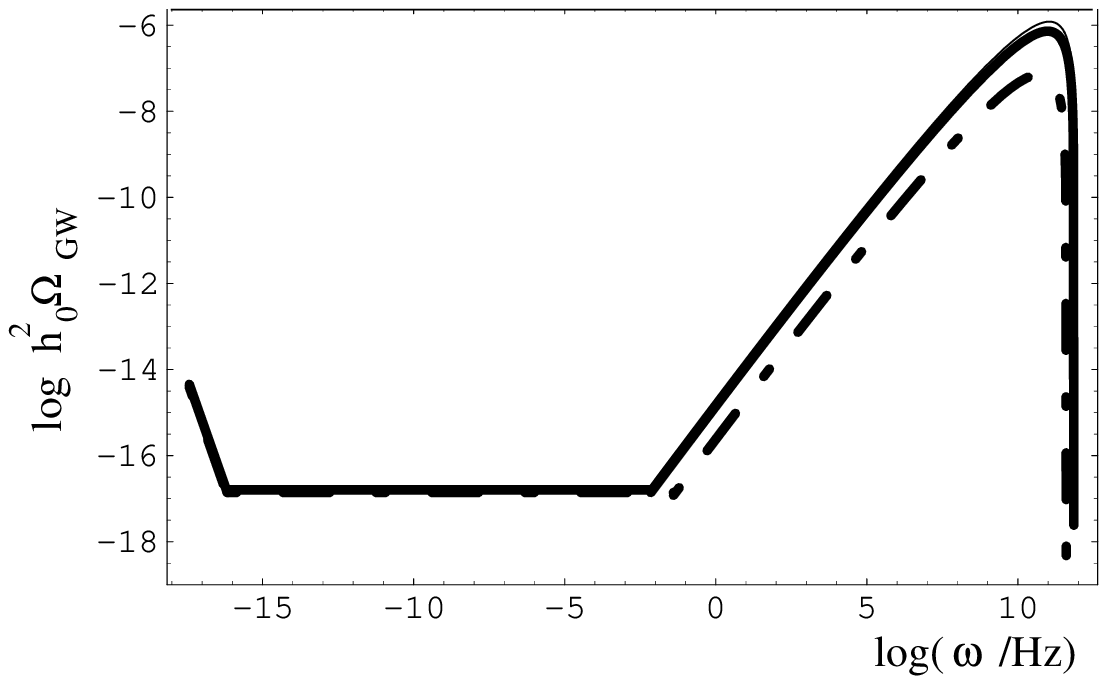}} &
      \hbox{\epsfxsize = 7 cm  \epsffile{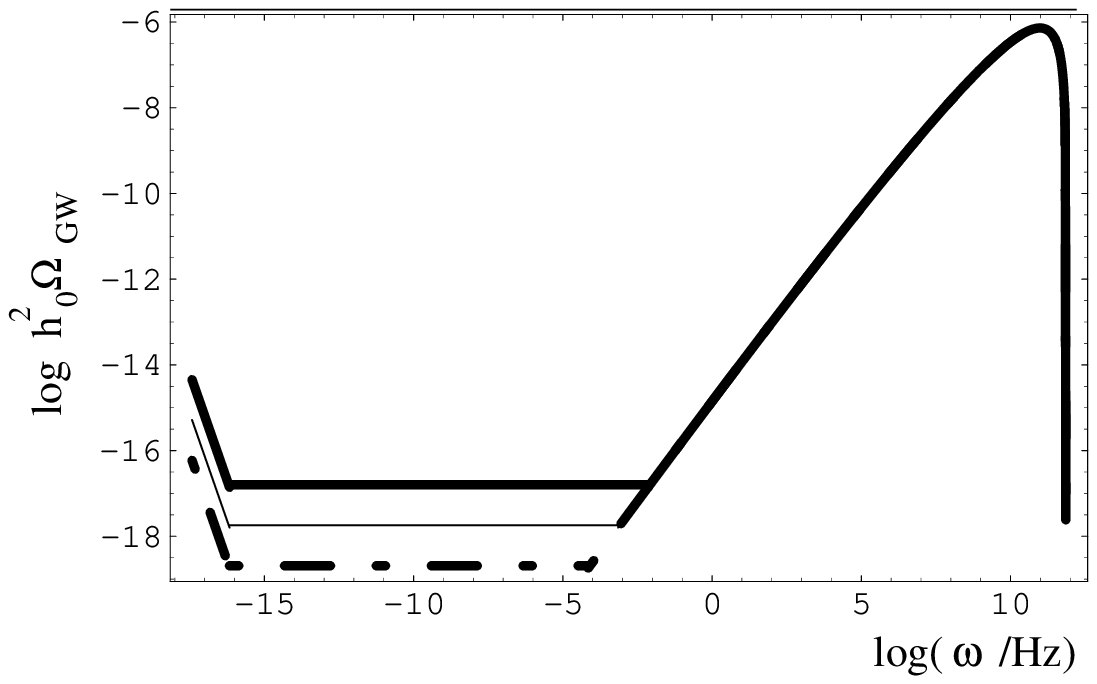}} \\
      \hline
\end{tabular}
\end{center}
\caption[a]{
In the left plot we report the energy spectrum  as a function of 
the physical wave number for a fixed value of $\lambda$ (
$\sim 10^{-7}$) but for different values of $R$. 
In particular we show the cases
$R= 10$ (dot-dashed line), the case $R=1$ (full thick line)
and the case $R=0.6$ (full thin line, almost invisible). 
We see that by {\em increasing} 
$R$ the height of the peak gets smaller and also its width shrinks. 
In the right plot 
we report the graviton energy spectrum for a fixed value of $R$ 
(which we take of order one) but for different values of $\lambda$ and, 
more specifically, $\lambda= 10^{-14}$ (full thick line), 
$\lambda= 10^{-15}$ (full thin line), $\lambda= 10^{-16}$ (dot-dashed line).
We took $\Omega_0 =0.3$, $g_{{\rm dec}}=3.36$ and 
$g_{{\rm th}}=106.75$ as fiducial
set of parameters. Notice that in spite of the fact  
that $\Omega_{{\rm GW}}(\omega,\eta_{0})$ does depend on the specific 
value of $h_0$ which appears, in our notations, in 
$\Omega_{\gamma}(\omega,\eta_0)$. We choose to plot 
$h^2_0 \Omega_{{\rm GW}}(\omega,\eta_0)$ which does not depend upon $h_{0}$.
In this way the {\em the amplitude} of the energy spectrum is independent on 
$h_0$ and the only quantities depending on $h_0$ are 
$\omega_0$ and $\omega_{{\rm dec}}$. In this plot
took $h_0=0.6$.}
\label{f4}
\end{figure}

An increase in $N_{s}$ does not only decreases the height of the peak,
 but it can also make the peak structure {\em narrower}. This happens 
simply because by {\em increasing } $N_{s}$, $\nu_{r} \propto 
N^{3/4}_{s}$ grows
and $\nu_1 \propto N^{-1/4}_{s}$ gets pushed towards more infra-red values. 
Given the limited range of variation of $R$ this effect is quite mild 
We illustrate the variation of $R$ on the 
energy spectrum in Fig. \ref{f4}. A decrease in the inflationary curvature 
scale  at the end of inflation 
{\em does not} affect the peak since the maximal 
amplified frequency does not depend on $H_1/M_P$ but 
it does only depends on $R$.
In quintessential 
inflationary models the energy density of relic gravitons
 can be much larger, at high 
frequencies, than in the case of ordinary inflationary models where the 
energy spectrum is still flat in the hard branch. The 
location of the peak is rather surprising. In fact it depends (very weakly,
as we said) on the number of minimally coupled scalar fields 
but {\em it does not depend upon the final curvature 
 scale at the end of inflation}. 
Thus the peak is firmly localized around $100$ GHz and it cannot move of 
one order of magnitude. This behavior has to be contrasted in ordinary 
inflationary models where the maximal frequency of the spectrum is 
determined by $10^{11}~\sqrt{H_1/M_{P}}$ Hz. So by lowering $H_{1}$, 
then, the maximal frequency decreases.
In the case we are discussing $H_1$ only appears in the expression of 
$\nu_{r}$. Therefore, by decreasing $H_1/M_{P}$ (see Fig. \ref{f4} 
right plot) $\nu_{1}$ does not move but $\nu_{r}$ gets comparatively smaller 
reducing the frequency range of the semi-hard branch.

\renewcommand{\theequation}{4.\arabic{equation}}
\setcounter{equation}{0}
\section{Gravitons Power Spectra and associated Spectral Densities} 
\begin{figure}
\begin{center}
\begin{tabular}{|c|c|}
      \hline
      \hbox{\epsfxsize = 7 cm  \epsffile{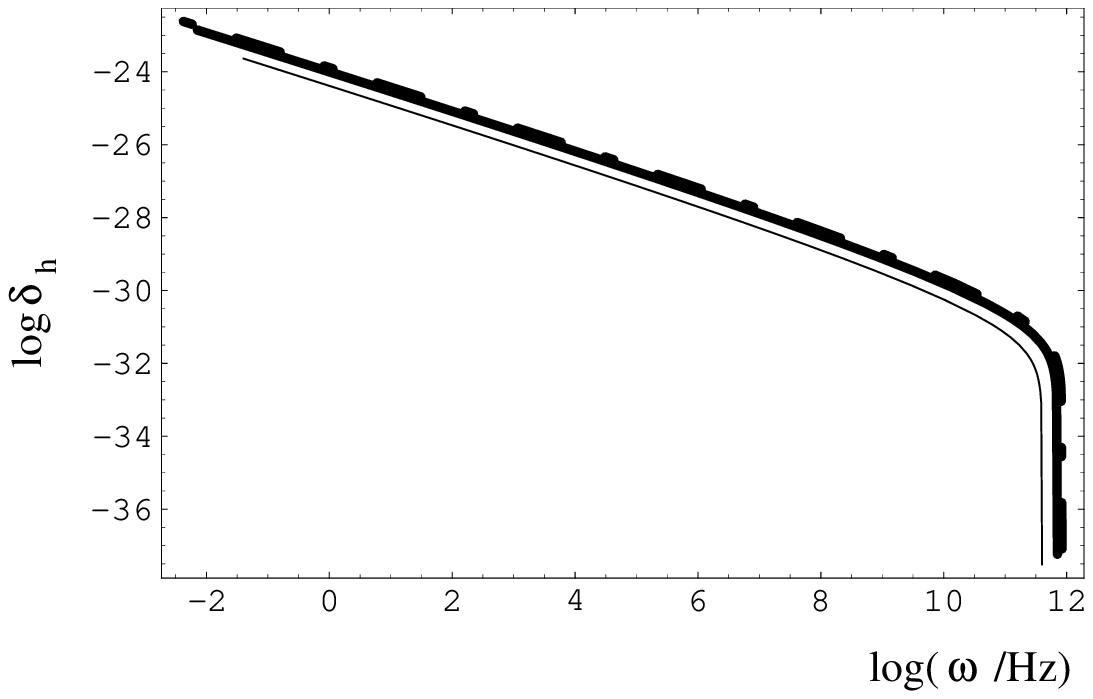}} &
      \hbox{\epsfxsize = 7 cm  \epsffile{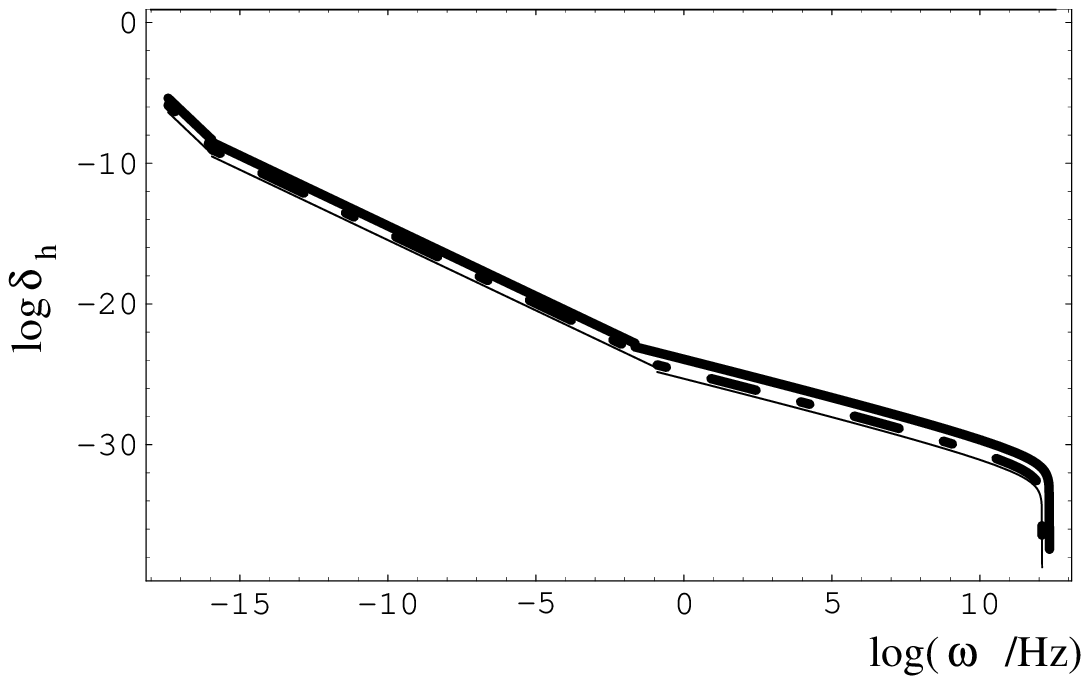}} \\
      \hline
\end{tabular}
\end{center}
\caption[a]{ At the left we plot  the hard branch of the 
power spectrum for a fixed value of
 $\lambda$ 
(which we take of order $10^{-14}$) but for different values of $R$ and, 
more specifically, $R=1$ (full thick line), 
$R=10$ (full thin line), $R=0.1$ (dot-dashed line).
 At the right 
we plot  the power spectrum for a fixed value of $R$ 
(which we take to be $10$) but for different values of $\lambda$ and, 
more specifically, $\lambda= 10^{-14}$ (full thick line), 
$\lambda= 10^{-15}$ (dot-dashed line), $\lambda= 10^{-16}$ (full thin line). 
In this and in the previous plot we took $h_0 =0.6$ and $\Omega_{0} =0.3$ as
 fiducial values. }
\label{f6}
\end{figure}

The relic graviton spectrum can be characterized not only in terms 
of the energy density but also in terms of the power spectrum, namely in terms 
of the Fourier transform of the two-points correlation function 
of the graviton field operators. Of course the energy density 
and the power spectrum can be precisely related. 
On the basis of the derivation reported in Appendix A 
(see Eq.  (\ref{endensspec})) we can connect the power spectrum
to the energy density of the relic graviton background 
\begin{equation}
\Omega_{{\rm GW}}(\omega, \eta_0) = \frac{1}{6\pi^2} 
\bigl(\frac{\omega}{H_0}\bigr)^2 |\delta_{h}(\omega,\eta_0))|^2,~~~~~
|\delta_{h}(k,\eta_0)|^2= k^{3} ( |h_{\otimes}(k,\eta_0)|^2 
+ |h_{\oplus}(k,\eta_0)|^2), 
\label{pp}
\end{equation}
where $h_{\otimes}(k,\eta)= 
\psi_{\otimes}(k,\eta)/a(\eta)$ and $h_{\oplus}(k,\eta)= 
\psi_{\oplus}(k,\eta)/a(\eta)$ are the 
Fourier amplitudes of the graviton field operators associated with the 
two (independent) polarizations [see Eqs. (\ref{expfield})] and $H_0$ is the 
present value of the Hubble parameter.
The three branches of our power spectrum turn out to be: 
\begin{eqnarray}
&&\delta_{h}(\omega,\eta_0) = {\cal B} ~R^{- \frac{3}{4}} 
\sqrt{\frac{3\varepsilon}{2\lambda}}~ \sqrt{\Omega_{\gamma}(\eta_0)} 
~\biggl(\frac{\omega}{\omega_{r}}\biggr)^{-\frac{1}{2}}~
\ln{\biggl(\frac{\omega}{\omega_1}\biggr)},~~~~~
\omega_{r}<\omega<\omega_{1}
\nonumber\\
&& \delta_{h}(\omega,\eta_0)= {\cal B}~R^{-\frac{3}{4}} 
\sqrt{\frac{3  \varepsilon~\pi}{8\lambda}}~ 
\sqrt{\Omega_{\gamma}(\eta_0)} ~\biggl(\frac{\omega}{\omega_r}\biggr)^{-1} 
~\ln{\biggl(\frac{\omega_{r}}{\omega_1}\biggr)}
,~~~~~\omega_{{\rm dec}}<\omega<\omega_{r}
\nonumber\\
&&\delta_{h}(\omega,\eta_0) = 
6.5\times 10^{-3}~~\frac{\pi\sqrt{6~\varepsilon~\lambda~\pi}}{8} ~
\sqrt{\frac{ \Omega_{\gamma}(\eta_0) }{\Omega_{0}(\eta_0)}}
\biggl(\frac{\omega_{{\rm dec}}}{\omega}\biggr)^2~
 \ln{\biggl(\frac{\omega_{r}}{\omega_1}\biggr)},~~~~~~~
\omega_{0}<\omega<\omega_{{\rm dec}}
\end{eqnarray}
where ${\cal B} = 3.07 \times 10^{-30}~h_0$.
Our results are illustrated in Fig. \ref{f6}. 
As we can see the power spectrum of the hard branch evolves 
typically as $\omega^{-1/2}$. Our power spectrum 
 declines slower than in 
ordinary inflationary models where the high frequency tail evolves typically 
as $\omega^{-1}$. This behavior occurs, in our case, for frequencies 
$\omega_{{\rm dec}}<\omega<\omega_{r}$. By varying $R$ in the allowed 
range the power spectrum is only slightly affected. We stress this point in 
Fig. \ref{f6} where different power spectra are reported for various 
values of $R$ at fixed lambda. 
A similar effect can be observed if we keep $R$ fixed and we let $\lambda$ 
free to change. Again the impact of the variation of $\lambda$ affects the 
power spectrum less than the energy density. The reason for this 
behavior is due to the fact that a variation in 
$\delta_{h}(\omega,\eta_{0})$ (at a given frequency) boils down to
a quadratic variation in $\Omega_{GW}(\omega,\eta_0)$.

In comparing the produced graviton spectrum with the experimental 
sensitivities of the various detectors it turns out to be 
useful to translate the physical information contained into the 
energy density into another quantity, the 
{\em spectral amplitude} (often called also {\em spectral density}) 
whose relation with the energy density has been derived in 
Appendix A:
\begin{equation}
S_{h}(\nu,\eta_0)= 8\times 10^{-37} ~
\Omega_{GW}(\nu,\eta_0) ~h_{0}^2~\frac{{\rm Hz}^2}{\nu^3}.
\label{spdens}
\end{equation}
From Eq. (\ref{spdens}) $S_{h}(\nu,\eta_0)$ 
 turns out to be quite 
a small fraction of a second. For this 
reason sometimes the experimentalists 
express their bounds in terms of $S^{1/2}_{h}(\nu,\eta_0)$. 
The maximal 
signal expected from a stochastic graviton background is limited 
(from above) by the nucleosynthesis bound. Let us then demand that 
the peak of the graviton spectrum does not exceed the 
nucleosynthesis bound. Then, to reach a level of sensitivity 
 comparable with $h_0^2 \Omega_{{\rm GW}}(\nu, \eta_0) \laq ~10^{-6}$
will imply that 
\begin{equation}
\sqrt{S_{h}(\nu,\eta_0)} \laq 3\times 10^{-26} 
\biggl( \frac{{\rm kHz}}{\nu}\biggr)^{3/2} {\rm Hz}^{-1/2}.
\end{equation}
The physical relevance of $S_{h}(\nu,\eta_0)$ is related to the way we hope 
to observe in the near future stochastic GW backgrounds.
In order to detect a 
gravitational wave background we in an optimal way \cite{gris2}
we need {\em at least} two detectors (two bars, 
two interferometers, one bar and one interferometer...). 
Suppose then that we
have two detectors and suppose that the output of a the detectors 
is given by $s_{i}= h_{i} + {\cal N}_i$ where $i=1,2$ refers to each single 
detector ; $h_{i}$ is the gravitational fluctuation to be detected and 
${\cal N}_{i}$ is the noise associated with each detector measurement. 
Now, if
the noises of the two detectors {\em are not correlated}, then, the ensemble 
average of the Fourier components of the noises is stochastic, namely
\begin{equation}
\langle {\cal N}^{\ast}_{i}(\nu) {\cal N}_{j}(\nu')\rangle = \frac{1}{2} 
\delta(\nu-\nu')\delta_{ij}S_{{\cal N}}(\nu),
\end{equation}
where $S_{{\cal N}}(\nu)$is the {\em spectral 
density of the noise}.
The noise level of the detector can then be estimated bt 
$\sqrt{S_{{\cal N}}(\nu)}$.
As we discussed in Appendix A (see Eqs. (\ref{specsens})--(\ref{specsens2})) 
it is also possible to characterize the 
{\em signal} with the same technique and, then, we will have 
\begin{equation}
\langle h_{i}(\nu) h_{j}^{\ast}(\nu') \rangle = 
\frac{1}{2}\delta_{ij}\delta(\nu-\nu') S_{h}(\nu).
\label{specsens3}
\end{equation}
where $S_{h}(\nu)= S_{h}(-\nu)$ 
is the {\em spectral density of the signal} and it is 
related to $\Omega_{{\rm GW}}$ by Eq. (\ref{spdens}). 
Very roughly,  if a signal is registered by a detector this will mean that 
$S_{h}(\nu_{\ast},\eta_0) > 
S_{{\cal N}(\nu_{\ast})}$, namely, the spectral density of the signal 
will be  larger, at a given frequency $\nu_{\ast}$,
 than the spectral density of the noise associated with the 
detector pair.  In order to 
confront our signal with the available sensitivities we need to compute the 
spectral density $S_{h}$. Taking into account the numerical factors, we get
\begin{eqnarray}
&&S_{h}(\omega,\eta_0) =  {\cal C}~ R^{- \frac{9}{4}}~
\bigl(\frac{g_{{\rm dec}}}{g_{r}}\bigr)^{-1}~\frac{\varepsilon}{\lambda^2}
 \Omega_{\gamma}(\eta_0)~ \biggl(\frac{\omega}{\omega_{r}}\biggr)^{-2}~
\ln^2{\biggl(\frac{\omega}{\omega_1}\biggr)}~
 {\rm Hz}^{-1},~~~~\omega_{r}<\omega<\omega_{1}
\nonumber\\
&&S_{h}(\omega,\eta_0) = \frac{{\cal C}\pi}{4}~ 
\bigl(\frac{g_{{\rm dec}}}{g_{r}}\bigr)^{-1}
~ R^{- \frac{9}{4}}~\frac{\varepsilon}{\lambda^2}
\Omega_{\gamma}(\eta_0)~ \biggl(\frac{\omega}{\omega_r}\biggr)^{-3} 
~\ln^2{\biggl(\frac{\omega_{r}}{\omega_1}\biggr)}~{\rm Hz}^{-1},~~~~~~
\omega_{{\rm dec}}<\omega<\omega_{r}
\nonumber\\
&&S_{h}(\omega,\eta_0) 
=4.2\times 10^{11}~~ 
\frac{ 3 ~\varepsilon~\lambda}{64\pi}~\Omega_{\gamma}(\eta_0)
 h_0^{-1}~\Omega_0^{-3/2} 
~ \biggl(\frac{\omega}{\omega_{{\rm dec}}}\biggr)^{-5} ~
\ln^2{\biggl(\frac{\omega_1}{\omega_r}\biggr)} ~{\rm Hz}^{-1},~~~~~
\omega_{0}<\omega<\omega_{{\rm dec}}
\end{eqnarray}
where ${\cal C} \sim~ 2.12~10^{-13}~{\cal B}^2$.

\begin{figure}
\centerline{\epsfxsize = 8 cm  \epsffile{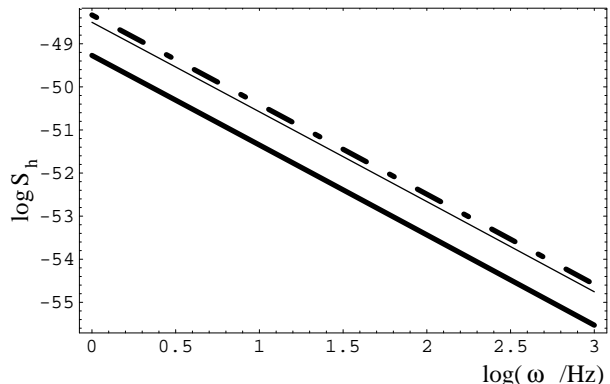}} 
\caption[a]{We plot the spectral density for $\lambda= 10^{-14}$ 
in the frequency 
range relevant for the forthcoming interferometric data on direct 
GW detection. In the full 
thick line we report the case $R=10$, in the full thin line the case $R=1$.
and in the dot-dashed line the case $R=0.6$. 
Again we took $h_0=0.6$ and $\Omega_0 =0.3$ as fiducial values. }
\label{f8}
\end{figure}

Not only the spectral 
amplitude of the theoretical signal depends upon the frequency, but also 
 spectral amplitude of the noise does depend upon the frequency. 
It is not only important if
 $S_{h}(\nu) > S_{{\cal N}}(\nu)$ at a particular frequency but it is also 
crucial to take into account, for detection strategies, 
{\em the spectral behavior of the signal} versus the 
{\em spectral behavior of the noise}  in the frequency range explored by 
the detectors. The  spectral density in the hard branch is illustrated
in  Fig. \ref{f8} from  $\omega= 1$ Hz until $\omega = 1$ KHz. We remind
that this  range of wave-numbers is the one relevant for the forthcoming 
interferometric detectors.
In Fig. \ref{f8} we see that the spectral density of our signal is 
mainly concentrated in the blank region between the 
thick line (corresponding to $R=10$) and the full thin line (corresponding 
to $R=1$). For $\omega \sim 0.1$ kHz, $S_{h}\sim 10^{-52}$--$10^{-53}$ sec.
For  $\omega \sim 0.01$ kHz, $S_{h}\sim 10^{-50}$--$10^{-51}$ sec. This 
observation  shows that, within the frequency range of the 
interferometers our theoretical signal can be larger or smaller 
depending upon the frequency. 
\renewcommand{\theequation}{5.\arabic{equation}}
\setcounter{equation}{0}
\section{Detectability of the Quintessential Graviton Spectra} 
There are, at the moment various interferometric detectors under 
construction. They include the two LIGO detectors \cite{ligo} being built
 by a joint Caltech/MIT
collaboration, the VIRGO detector (near Pisa, Italy) \cite{virgo} 
the GEO-600 (Hannover, Germany) \cite{geo} 
and the TAMA-300 (near Tokyo, Japan) \cite{tama}. 
The noise spectral densities of these detectors, defined in a frequency 
range going from $1$ Hz to $10^{4}$ Hz,  decline 
usually quite rapidly from $1$ to $100$ Hz, they have a minimum 
(around $100$ Hz) corresponding to the maximal sensitivity and then they rise 
again with a more gentle slope until, approximately, $1$--$10$ kHz. 
\begin{figure}[htb!]
\centerline{\epsfig{file=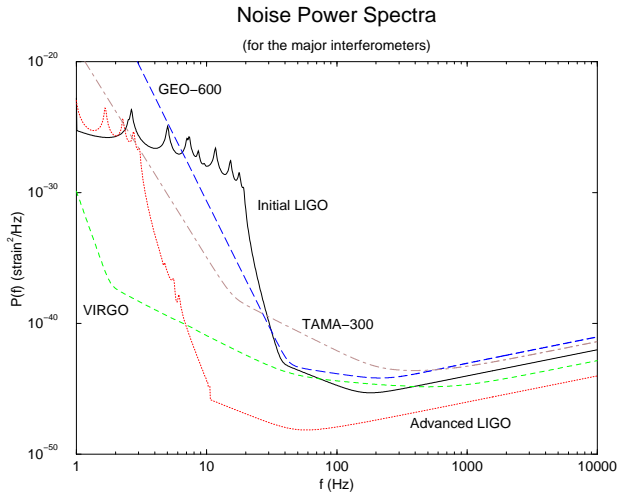,angle=-90,width=3.4in,bbllx=25pt,bblly=50pt,bburx=590pt,bbury=740pt}}
\caption[a]{\label{f9}
The predicetd noise power spectra for various interferometers. This Figure is 
adapted from \cite{alrom}. $P(f)$ is what we called $S_{\cal N}$ and it is the 
quantity which should be compared with $S_{h}$. In this plot $f\equiv \nu$.}
\end{figure}
As we discussed in the previous Section, in order to have some 
hopes of detection we have to demand that the theoretical spectral density 
of the signal is larger than the spectral density of the noise. 
Let us then try to compare this value with the expected 
sensitivity of the interferometers possibly available in the 
near future. The power spectrum of our signal is  too 
small to be seen by TAMA-300, GEO-600, and VIRGO.  
By correlating different detectors the sensitivity can 
increase also  by a large factor \cite{M,C,F}. However, the published 
results on the foreseen sensitivities at $\nu_{I}$ are far too large 
to be relevant for our background  \cite{sph}.
By comparing Fig. \ref{f8} with Fig. \ref{f9} we can argue that only the 
advanced LIGO detectors are closer to our predicted spectral density and that 
our signal in generally smaller than the advanced LIGO sensitivity.
The  two (identical) LIGO detectors 
are under construction in Handford (Washington) and in Livingston (Lousiana).
After various years of operation the detectors will be continuously 
upgraded reaching, hopefully, 
the so-called {\em advanced} level of sensitivity.

Let us estimate the strength of our background for a frequency of the 
order of $0.1$ kHz --$1$ kHz. 
Let us assume that the energy density of the 
stochastic background is the maximal compatible with the nucleosynthesis 
indications. 
The graviton energy density (in critical units)
 at a frequency $\nu_{I}\sim 0.1$--$1$ kHz is then 
\begin{eqnarray}
&&\Omega_{{\rm GW}}(\nu_{I}, \eta_0)~h_{0}^2
= 2.29 ~10^{-15}~ N_{s}^{-3/4}~ \bigl[ -19.7 + 0.25~
\ln{N_s}\bigr]^2,~~~~~\nu_{I} = 0.1~ {\rm kHz}, 
\label{sens1}\\
&& \Omega_{{\rm GW}}(\nu_{I}, \eta_0)~h_{0}^2
= 2.29~10^{-14}~N_{s}^{-3/4}~\bigl[ -17.4 + 0.25~ 
\ln{N_{s}}\bigr]^2,~~~~\nu_{I} = 1~{\rm kHz}.
\label{sens2}
\end{eqnarray}
Suppose then that we 
correlate the two LIGO detectors for a period $\tau = 4$ months. Then, 
the signal to noise ratio (squared) can be expressed as \cite{M,C,F}
\begin{equation}
\biggl( \frac{S}{N}\biggr)^2  = \frac{9 H_{0}^4}{50 \pi^4} \tau 
\int_{0}^{\infty} d\nu \frac{\gamma^2(\nu) \Omega^2_{{\rm GW}}(\nu,\eta_0)}
{\nu^6 S^{(1)}_{{\cal N}}(\nu)S^{(2)}_{{\cal N}}(\nu)}.
\label{SNR}
\end{equation}
The function $\gamma (\nu)$ is called the overlap function. It takes
into account the difference in location and orientation of the two
detectors. It has been computed for the various pairs of LIGO-WA, LIGO-LA,
VIRGO and GEO-600 detectors~\cite{F}. For detectors very close and parallel,
$\gamma (\nu)=1$. Basically, $\gamma (\nu)$
  cuts off the integrand of Eq. (\ref{SNR}) 
at a frequency $2\pi \nu$ of the order of the inverse separation between
the two detectors. For the two LIGO detectors, this cutoff is around
60 Hz. $S^{(1,2)}_{{\cal N}}$ are the noise spectral densities of the 
two LIGO detectors and since the two detectors are supposed to be 
identical we will have that $S^{(1)}_{{\cal N}}(\nu)=
S^{(2)}_{{\cal N}}(\nu)$. In order to detect a stochastic background with 
$90\%$ confidence we have to demand $S/N \gaq 1.65$.
Now in order to estimate the signal to noise ratio we need to estimate 
numerically the integral appearing in Eq. (\ref{SNR}). We know the theoretical 
spectrum since we just estimated it.
The noise spectral densities  of the LIGO detectors are not of 
public availability
 so that we cannot perform numerically this integral. 
In the case of a flat energy spectrum we have that the minimum 
$\Omega_{{\rm GW}}$ detectable in $\tau= 4$ months is given, with $90~\%$ 
confidence, by $\Omega_{{\rm GW}}(\nu_{I},\eta_{0})= 
5\times 10^{-6} h^{-2}_{0}$ (for the 
initial LIGO detectors) and by
 $\Omega_{{\rm GW}}(\nu,\eta_0)
= 5\times 10^{-11} h^{-2}_{0}$ (for the advanced LIGO
detectors). For a correct comparison we should not confront Eqs. (\ref{sens1})
and (\ref{sens2}) with the sensitivities of the LIGO detectors to flat 
energy spectra but rather with the sensitivities obtained from Eq. (\ref{SNR})
in the case of our specific energy spectrum reported in 
Eqs. (\ref{gravendens}).

Let us then compare, for illustration Eqs. (\ref{sens1}) and (\ref{sens2}) with
the sensitivity to a flat energy spectrum even if this is not 
completely correct. The idea is to discuss, at fixed frequencies, the 
maximal signal provided by Eqs. (\ref{sens1})--(\ref{sens2}) 
for different values of $R$. 
\begin{figure}
\begin{center}
\centerline{\epsfxsize = 8 cm  \epsffile{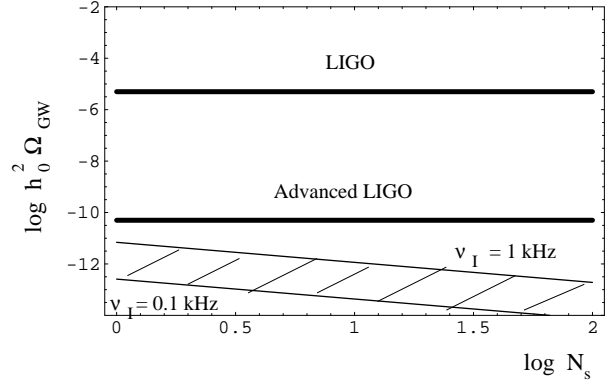}} 
\end{center}
\caption[a]{ With the two full thin lines 
we  illustrate $\log{h^2_{0}\Omega_{{\rm GW}}(\nu_{I},\eta_0)}$ 
as a function of $\log{N_{s}}$ for frequencies 
$\nu_{I}= 0.1 $ kHz according to Eqs. (\ref{sens1})--(\ref{sens2}). 
The full thick lines 
represents the sensitivity of the LIGO detectors 
and of the advanced LIGO detectors to an energy 
density with flat frequency spectrum. From top to bottom the 
thick lines refer, respectively, to $h^2_0 
\Omega_{{\rm GW}} =5 \times 10^{-6}$ and to 
$h^2_0 
\Omega_{{\rm GW}} =5 \times 10^{-11}$. In order to be detected, the 
theoretical signal has to be above the thick line(s). We see that the region 
between the two thin lines does not overlap with the sensitivity of the 
upgraded LIGO detectors by, roughly, 1.5 orders of magnitude.}
\label{f10}
\end{figure}
This comparison is illustrated in Fig. \ref{f10}. 
For the allowed range of variation of $N_{s}$
our signal lies always below (of roughly $1.5$ orders of magnitude)
the predicted sensitivity 
for the detection, by the advanced LIGO detectors, of
 an energy density with flat slope. The main 
uncertainty in this analysis is however the spectral behavior of the 
sensitivity for a spectrum which, unlike the one used for comparison,
is not flat. It might be quite interesting to perform this calculation 
in order to see which is the precise sensitivity of the LIGO 
detectors to a spectral energy density as large as $10^{-12}$ and  
rising as $(\nu /\nu_{r})\ln{(\nu /\nu_1)}$ in a frequency range 
$1$ Hz--$1$ kHz. 

If we move from the kHz region to the GHz region the signal gets much larger 
than in the ordinary 
inflationary models. Moreover, the energy spectrum exhibits a quite 
broad peak whose typical amplitude can be as large as (but smaller then) 
$0.5 \times ~10^{-5}$. The spike (corresponding to the 
maximum of the peak) is located at a frequency of
 the order
 of $350~ R^{-1/4}$ GHz. 

On top of  the interferometric detectors there are other 
types of detectors which are operating at the moment or might be operating 
in the future. The available upper limits on the strength of a stochastic 
background of relic gravitons come from the (cryogenic) resonant-mass 
detectors. In particular EXPLORER, while operating at CERN, 
provided an upper (direct) bound on the graviton energy density. The 
frequency $\nu_{B}$ relevant for this type of detectors is among 
$907$ and $923 $Hz. By analyzing the 1991--1994 data of the  EXPLORER 
antenna, the Rome group of G. Pizzella got a bound on the strength of the 
gravitons energy density of the order of $h_{0}^2 \Omega_{{\rm GW}} \laq 
300$. This bound, as it is, is not crucial since we know that the graviton 
energy density should be much lower, however we stress that this is direct 
bound from an operating device. Moreover, by correlating two bar detectors 
with the same features (like EXPLORER, NAUTILUS or AURIGA
\footnote{ NAUTILUS (in Frascati, Rome) and AURIGA \cite{cer} 
(in Legnaro, Padova, Italy) are both resonant mass detectors.}) 
it is not excluded
that a very interesting sensitivity of $h_0^2\Omega_{{\rm GW}} 
\sim 10^{-3}$ will be reached.

The particular spectral shape of the signal coming from quintessential 
inflation  seems to point towards the use of electromagnetic detectors and, in 
particular of microwave cavities. 
A typical signature of the background we are discussing 
in the present paper is that the peak frequency (which almost 
saturates the nucleosynthesis bound) occurs for frequencies of the 
order of $\nu_{1} = 3.58\times 10^{11} R^{-1/4} $ Hz. 
We can say that, very roughly, the size of the GW 
detectors is not only determined by construction requirement but 
also by the typical frequency range of the spectrum we ought to explore. 
In this sense the large separation between the two LIGO detectors 
is connected with the fact that the explored  frequency range is of the 
order of $100$ Hz. Thus, if we deal with frequencies which are of the order 
of the GHz we can expect small detectors to be, theoretically, 
a viable option. 

Microwave cavities can be used as GW detectors in the 
 GHz frequency range \cite{picasso,caves}.
These detectors consist of an electromagnetic resonator, with two levels whose 
frequencies $\nu_{s}$ and $\nu_{a}$ are both much larger than the frequency 
$\nu_{{\rm GW}}$ of the gravitational wave to be detected. 
In the case of \cite{picasso} 
the two levels are achieved by coupling two resonators one symmetric in the 
electric fields and the other antisymmetric. Indeed, 
in the case of cylindrical microwave cavities  there are different normal 
oscillations of the electric fields. In \cite{picasso} the relevant 
mode for the experimental apparatus is the ${\rm TE}_{011}$  according to 
the terminology usually employed in electrodynamics 
in order to identify normal modes of a cavity corresponding to different 
boundary conditions  \cite{jackson}.
There were published results reporting the construction of such 
a detector \cite{melissinos}. In this case $\nu_{{\rm GW}} = 10 $ GHz
and $\Delta \nu = \nu_{s} - \nu_{a} \sim 1 $ MHz. In this experiment 
a sensitivity of fractional deformations 
$\delta x/x$ of the order of $10^{-17}$ was observed using an integration 
time $\Delta t \sim 10^{3} $ sec. 
The sensitivity to fractional deformations can be connected to the sensitivity 
for the observation of a monochromatic gravitational wave 
of frequency $\nu_{{\rm GW}}$. Following \cite{melissinos} we can learn 
that
 the sensitivity to fractional deformations is a function of
$P_{a}$ and $P_{s}$ ( the powers stored in the symmetric and 
antisymmetric levels), $Q$ (the quality factor of the cavity \cite{jackson}
and which gives 
 rate of dissipation of the power stored in the cavity). 
If we would assume, as in \cite{melissinos} 
$Q\sim 10^{9}$, $P_{a}\sim 10^{-21} $ Watt 
$P_{s}= 2.4\times 10^{-2}$ Watt , we would get $\delta h\sim 10^{-17}$.

There, are at the 
moment, no operating prototypes of these detectors and so it is difficult 
to evaluate their sensitivity. The example we quoted 
\cite{melissinos} refers to 1978. We think that possible improvements 
in the $Q$ factors can be envisaged (we see quoted values of the order of
$10^{12}$ which would definitely represent a step forward for the sensitivity).
In spite of the fact that improvements can be foreseen we can 
notice immediately that, perhaps, to look in the highest 
possible frequency range of our model is not the best thing to do. In fact 
from Eq. (\ref{pp}) we can argue that in order to detect a signal of the order
of $h^2_{0}\Omega_{{\rm GW}}\sim 10^{-6}$ at a frequency of $1$ GHz, we would 
need a sensitivity of the order of $\delta h\sim 10^{-30}$. Moreover  
as stressed in \cite{th} the thermal noise should be properly 
taken into account in the 
analysis of the outcome of these microwave detectors. Indeed, as noticed 
from the very beginning \cite{melissinos}, the thermal noise 
is one of the fundamental source of limitation of the sensitivity.
An interesting strategy could be to decrease 
 the operating frequency  range of the device 
by going at frequencies of the order of $1$ MHz.
Based on the considerations of \cite{picasso}
we can say that by taking high quality resonators the 
foreseen sensitivity can be as large as $h^2_{0}\Omega_{{\rm GW}}
 \sim 10^{-4}$. This sensitivity, though still above our signal, 
would be  quite promising.

\renewcommand{\theequation}{6.\arabic{equation}}
\setcounter{equation}{0}
\section{Graviton Spectra for quasi-de Sitter Phases} 

During an inflationary phase the evolution is not exactly of de Sitter type.
We want to understand how the slope of the energy density of the relic graviton
background will be modified by the slow-rolling corrections for frequencies 
accessible to the forthcoming interferometers.  
In general the deviations from a de Sitter stage can be induced either 
because of the specific inflationary model or because of the slow-rolling 
corrections whose strength can be described in terms of the so-called
slow-rolling parameters 
\begin{equation}
\alpha = - \frac{\dot{H}}{H^2}<~1,~~~
\beta = \frac{\ddot{\phi}}{H\dot{\phi}}<~1,
\label{sloropar}
\end{equation}
where $H = (\ln{a})^{\cdot}$ is the Hubble parameter in cosmic time,
$\phi$ is the inflaton  and the dot denotes derivation 
with respect to cosmic time.
In the slow rolling approximation the inflaton evolution is dominated by the 
scalar field potential according to the (approximate) equations
\begin{equation}
3 H \dot{\phi} + \frac{\partial V}{\partial \phi} \simeq 0, ~~~
M_{P}^2 H^2 \simeq V.
\label{sloroeq}
\end{equation}
In this
 approximation $H$ is not exactly constant but it slowly decreases 
leading to what we called quasi de Sitter phase.
The evolution equation of the mode function can be written, by using 
the definition of $\alpha$ as 
\begin{equation}
\psi'' + \bigl[ k^2 - 2~a^2~H^2\bigl(1 - \frac{\alpha}{2}\bigr) \bigr]\psi =0.
\label{sloromode}
\end{equation}
As we can see, the time-dependent frequency appearing in the mode function 
contains two contribution: the first one ( i.e. $2 a^2 H^2 = 2/\eta^2$) is 
the term coming from a pure de Sitter phase, the second one, proportional 
to $\alpha$, is the correction. 

In short the logic is the 
following. The quasi de Sitter phase modifies (through and $\alpha$-dependent 
correction)  the index of the Hankel functions whose precise value ( 
equal to $1.5$ in the pure de Sitter phase) gets slightly smaller than $1.5$.
From the sign of the correction appearing in Eq.
 (\ref{sloromode}) we can argue that the quasi de Sitter nature of the 
inflationary phase will lead to a  decrease in the slope of the energy 
spectrum. The question is how much the slope of the hard branch will 
be affected, or, more precisely, how much smaller than one will it be 
the slope of the hard branch of the graviton energy density. 

The answer to this question will of course depend upon the 
specific inflationary model since the size of the slow-rolling 
corrections can vary from one inflationary potential to the other.
Notice that our concern is
 different from the one usually present 
\cite{kolb,st,rep,rev2} in the context of  ordinary inflationary 
models where the slow-rolling 
corrections are taken into account in 
the soft branch of 
the spectra (namely at sufficiently large scales). Indeed, 
for flat spectra the most significant bounds come from the infra-red, whereas, 
in our case the most significant bounds 
are in the ultraviolet and we have
 to understand how the scales which re-enter in the stiff
 phase are affected by 
the quasi-de Sitter nature of the inflationary phase.
Of course, we could also discuss the slow-rolling coprrections to the 
soft branch of our spectrum, but, they are, comparatively, 
less relevant (for the structure of the spike) than the correction 
to the hard branch.
 
As usual, the slow-rolling corrections 
not only affect the mode function evolution but also the definition 
of conformal time itself, namely we will have 
\begin{equation}
\eta = \int \frac{dt}{a}= \int \frac{ da}{ a^2 H} = -\frac{1}{a H} + 
\int \alpha \frac{ da }{ a^2 H}.
\end{equation}
By using the fact that $ M_{P}^2 \dot{H} = - (3/2) \dot{\phi}^2$ and the fact 
that $\beta<1$
we can connect directly $\alpha$ to the slope of the potential 
\begin{equation}
\alpha = - \frac{\dot{H}}{H} = \frac{M_{P}^2}{6} 
\biggl( \frac{\partial \ln{V}}{\partial\phi}\biggr)^2.
\label{defa}
\end{equation}
Since we are interested in the lowest order slow-rolling correction 
we will assume that $\alpha$ and $\beta$ are constants. This 
is a simplification which will not affect (numerically ) the slope of the 
spectrum. In the case of inflationary models with chaotic and 
exponential potential  it can be shown that the ``running'' of $\alpha$ with 
$\phi$ (and, therefore, with the wavenumber $k$) will affect the spectral 
slopes with a term which is of the order of $1/N^2$ where $N$ is the number 
of inflationary e-folds \cite{rev2,turner1,turner2}. 
If $\alpha$  is constant then we have
 that $ \eta^{-1} = - (a H)/( 1 + \alpha)$ which leads, once inserted in Eq. 
(\ref{sloromode}) 
\begin{equation}
\psi'' + \bigl[ k^2 - \frac{\nu^2 - \frac{1}{4}}{\eta^2}\bigr] =0,~~~ 
{\rm with} 
~~~\nu = \frac{3}{2} + \alpha,
\end{equation}
where the expression of $\nu$ holds for $\alpha <1$. 

Let us examine now the quasi-de Sitter nature of different inflationary
scenarios.
An inflationary potential of chaotic form 
\begin{equation}
V(\phi) = \frac{ \phi^n}{n !},
\label{pot1}
\end{equation}
will lead to $ \alpha = (M_{P}^2/6) ( n^2/\phi^2)$. Let us take the 
value of $\alpha$ the one corresponding to modes crossing  the horizon
around $20$--$25$ e-folds before the end of inflation. 
The reason for this choice is very simple. We want to understand how is 
modified the slope of the energy spectrum for scales larger than 
(and of the order of)  
the ones probed by the interferometers. A simple calculation shows 
that, for instance, the LIGO/VIRGO scale 
crosses the horizon roughly $21$ e-folds before the end of  inflation.
The end of inflation 
occurs when $\alpha (\phi_{\rm end}) =1$.  Thus, at the end of 
inflation $\phi^2_{{\rm  end}}= (M_{P}^2/6) n^2$. By solving  consistently 
Eqs. (\ref{sloroeq}) with the potential (\ref{pot1}) we get 
easily that the number of inflationary e-folds at a given value of 
$\phi$ is given by
\begin{equation}
N(\phi) = \int_{\phi}^{\phi_{{\rm end}}} \frac{d a }{a} = 
\frac{3}{2 M_{P}^2 n} ( \phi^2 - \phi^2_{{\rm end}}).
\end{equation}
From this last equation we can determine easily $\phi_{21}$ and then, by
inserting $\phi^2_{{\rm end}}$, $\alpha$ turns out to be
\begin{equation}
\alpha = \frac{n}{n + 84},
\end{equation}
which is of the order of $0.02$ for $n=2$ , of the order of 
$0.04$ for $n=4$ and so on. 
Another example could be the one 
of an exponential potential. 
Using the definition of $\alpha$ from Eq. (\ref{defa}) we have that 
for an exponential potential of the form 
\begin{equation}
V(\phi) = \exp{\bigl[ \frac{1}{q} \frac{\phi}{M_{P}}\bigr]},
\end{equation}
will lead to $\alpha = 1/q^2$.

With these results we can easily compute the corrections to the 
hard branch of the spectrum. The spectral energy density will be 
\begin{eqnarray}
&&\Omega_{{\rm GW}}(\omega,\eta_0) = \Omega_{\gamma}(\eta_0) 
f( \alpha, \lambda, R) 
\bigl(\frac{\omega}{\omega_{r}}\bigr)^{1 - 2 \alpha} 
\ln^2{\bigl(\frac{\omega}{\omega_1}\bigr)},~~~{\rm with}
\nonumber\\ 
&&f(\alpha, \lambda, R) =\frac{9}{\pi^4} ~\bigl( \alpha + \frac{3}{2}\bigr)^2~ 
2^{2 \alpha} ~|\Gamma(\frac{3}{2} + \alpha)|^2 \lambda^{ 1 - 2 \alpha} 
R^{- 2\alpha} \bigl(\frac{g_{{\rm dec}}}{g_{r}}\bigr)^{1/3}.
\label{corr}
\end{eqnarray}
The power spectrum and the associated spectral density  can be computed 
from using the techniques already discussed in the previous Sections.
Some authors call $-2\alpha$ the corrections to the spectral index. 
Eq. (\ref{corr}) tells 
us that the corrections arising during a quasi-de Sitter phase 
always go in the direction of making the maximal slope of the hard branch 
slightly smaller than one by a factor $0.08$ (for instance in the case
 the case of $\phi^4$ potential). 
In the case of exponential potential the magnitude of the
 correction to the slope 
is controlled by $q^2$. Again we see that for reasonable values of $q^2$
 like 
$q^2=5,~10,...$ \cite{rev} the correction are again small and, at 
most, of the order of few percents.

\renewcommand{\theequation}{7.\arabic{equation}}
\setcounter{equation}{0}
\section{Concluding Remarks} 

In this paper we computed the relic graviton spectra in quintessential 
inflationary models. We showed that the energy spectra possess
a hard branch which increases in frequency.
Large energy densities in relic gravitons can be expected 
in this class of models. The  spike of the energy spectrum can be as large 
as $h_{0}^2 \Omega_{{\rm GW}} \simeq 10^{-6}$ at a typical (present)
frequency $\nu_{1} = 358\times R^{-1/4}$ GHz.
The spectral amplitude at the interferometers 
frequencies is just below the value
visible by the upgraded LIGO detectors. 
Since a large amount of  energy  density can be stored around 
the GHz the use of small electromagnetic 
detectors for the detection of such a background seems more 
plausible. In particular 
microwave cavities should be considered as a possible candidate. 
Our investigation also hints that the sensitivities of the advanced 
LIGO detectors to energy spectra increasing, in frequency,  as
 $\nu\ln\nu$ should be precisely computed by convolving our 
spectra with the noise power spectra of the detectors. 

\section*{Acknowledgments}
I would like to thank Alex Vilenkin for very useful 
discussions and comments which stimulated the present investigation. 
I thank E. Coccia, E. Picasso and G. Pizzella for various 
informative discussions. I would also like to thank D. Babusci and 
S. Bellucci for a kind invitation in Laboratori Nazionali di Frascati
(LNF) where some of the mentioned discussions took place.
\newpage
\begin{appendix}

\renewcommand{\theequation}{A.\arabic{equation}}
\setcounter{equation}{0}
\section{Relic Gravitons Correlation Functions} 

In this Appendix we concentrate some of the more technical derivations 
concerning the spectral properties of the relic gravitons which 
can be characterized through their energy density in critical units
 or through the two point correlation
function of the amplified tensor fluctuations of the metric. 
We want to give explicit derivations of  these
quantities and of their relations.

Let us start from the canonical action describing relic gravitons in 
a curved metric of Friedmann-Robertson-Walker type. GW, 
being pure tensor modes of the geometry, only couple to the curvature 
but not to the matter sources. Their effective action can be obtained 
by perturbing the Einstein-Hilbert action
\begin{equation}
S = -\frac{1}{6 l_{P}^2} \int \sqrt{-g} R d^4 x,
\end{equation}
to second order in the amplitude of the tensor fluctuations of the metric
\begin{equation}
g_{\mu\nu}(\vec{x},\eta) \rightarrow \overline{g}_{\mu\nu}(\eta) +
 \delta g^{(T)}_{\mu\nu}(\vec{x},\eta),~~~
\delta g_{\mu\nu} = h_{\mu\nu}, ~~~h_{\mu 0} =0.
\end{equation}
Pure tensor fluctuations of the metric 
(with two physical polarizations) can be constructed by using a
 symmetric three-tensor $h_{ij}$ satisfying the constraints $h_{i}^{i}=0$, 
$\nabla_{i} h_{j}^{i}=0$ (where $\nabla_{i}$ is the covariant derivative 
with respect to the three-dimensional background metric).
The action perturbed to second order in the amplitude of the 
tensor fluctuations becomes
\begin{equation}
\delta^{(2)}S^{(T)} = \frac{1}{24 l_{P}^2} \int d^4 x \sqrt{- \overline{g}} 
\partial_{\alpha} h_{ij} \partial_{\beta} h^{ij} \overline{g}^{\alpha\beta}.
\end{equation}
We can construct two independent scalars corresponding to the two 
polarization states of the gravitational  waves:
\begin{equation}
h_{ij}(\vec{x},\eta)= h_{\oplus}(\vec{x},\eta) q_{ij}^{(1)}(\vec{x}) + 
h_{\otimes}(\vec{x},\eta) q_{ij}^{(2)}(\vec{x}),
\end{equation}
where 
\begin{equation}
q_{(1)}^{ij}(\vec{x}) = - [ e_{1}^{i}(\vec{x}) e^{j}_{1}(\vec{x}) - 
  e_{2}^{j}(\vec{x}) e^{i}_{2}(\vec{x})],~~~
q_{(2)}^{ij}(\vec{x}) = - [ e_{1}^{i}(\vec{x}) e^{j}_{1}(\vec{x}) + 
  e_{2}^{j}(\vec{x}) e^{i}_{2}(\vec{x})].
\end{equation}
In these coordinates $\vec{e}_{1}$, $\vec{e}_{2}$ and $\vec{e}_{3}$ form an
orthonormal set of vectors. In the previous decomposition we assumed 
the wave to propagate along $\vec{e}_{3}$. So the perturbed action can also 
be written, in units of $6 l_{P}^2=1$, as 
\begin{equation}
\delta^{(2)} S^{(T)} = \frac{1}{2} \int d^{4} x  
\sqrt{- \overline{g}} [ \partial_{\alpha} h_{\oplus} 
\partial_{\beta}h_{\oplus} \overline{g}^{\alpha\beta} +
\partial_{\alpha} h_{\otimes} 
\partial_{\beta}h_{\otimes} \overline{g}^{\alpha\beta} ].
\end{equation}
The associated energy-momentum tensor can be written as \cite{ford2,ford3}
\begin{equation}
T_{\mu\nu} = \frac{1}{2} \biggl[ 
\partial_{\mu} h_{\oplus} \partial_{\nu} h_{\oplus}
 + \partial_{\mu} h_{\otimes} \partial_{\nu} h_{\otimes} -
 \overline{g}_{\mu\nu} 
\biggl( \overline{g}^{\alpha\beta}
\partial_{\alpha}h_{\oplus} \partial_{\beta} h_{\oplus}
+ \overline{g}^{\alpha\beta}\partial_{\alpha}h_{\otimes} 
\partial_{\beta} h_{\otimes}\biggr)\biggr].
\end{equation}
The canonically normalized graviton field can 
be written as $\mu_{ij} = a h_{ij}$ and its corresponding action reads
\begin{equation}
\delta^{(2)} S^{(T)} = \frac{1}{2} \int d^3 x d\eta \biggl[ \eta^{\alpha\beta}
 \partial_{\alpha}\mu_{\oplus}\partial_{\beta}\mu_{\oplus} + \eta^{\alpha\beta}
 \partial_{\alpha}\mu_{\otimes}\partial_{\beta}\mu_{\otimes} + \frac{a''}{a}( 
\mu_{\oplus}^2 + \mu_{\otimes}^2) \biggr].
\end{equation}
With this form of the action we can promote the classical
 canonical normal modes to 
field operators obeying the standard (equal time) commutation relations. 
The Fourier expansion of the field operators will then look
\begin{eqnarray}
&&\hat{\mu}_{\oplus}(\vec{x},\eta) = \frac{1}{(2\pi)^{3/2}} \int d^3 k \biggl[
 \hat{\mu}_{\oplus}(k,\eta) e^{i \vec{k}\cdot\vec{x}} +  
\hat{\mu}_{\oplus}^{\dagger}(k,\eta)  e^{-i \vec{k}\cdot\vec{x}}\biggr],
\nonumber\\
&&\hat{\mu}_{\otimes}(\vec{x},\eta) = \frac{1}{(2\pi)^{3/2}} \int d^3 k \biggl[
 \hat{\mu}_{\otimes}(k,\eta) e^{i \vec{k}\cdot\vec{x}} +  
\hat{\mu}_{\otimes}^{\dagger}(k,\eta)  e^{-i \vec{k}\cdot\vec{x}}\biggr],
\label{expfield}
\end{eqnarray}
where $ \hat{\mu}_{\oplus}(k,\eta) = 
\psi_{\oplus}(k,\eta) \hat{a}_{\oplus}(\vec{k})$
 and 
 $ \hat{\mu}_{\otimes}(k,\eta) = \psi_{\otimes}(k,\eta)
 \hat{a}_{\otimes}(\vec{k})$.
The creation and annihilation operators follow the usual commutation
 relations namely
\begin{equation}
[\hat{a}_{\lambda}(\vec{k}), \hat{a}_{\sigma}^{\dagger}(\vec{p})] = 
\delta_{\lambda\sigma} \delta^{(3)}(\vec{k}- \vec{p}).
\end{equation}
From the Heisenberg equations of motion for the field operators
 follow the evolution 
equations for the Fourier amplitudes
\begin{equation}
\psi'' + [k^2  - \frac{a''}{a} ] \psi =0,
\label{psi}
\end{equation}
for each of the two physical polarizations. 
In most of the cosmological applications we will be dealing 
with $|a''/a|$ has a bell-like shape going to zero , for 
large conformal times, as $\eta^{-2}$.
Eq. (\ref{psi}) can be be solved in two significant limits. 
The first one is for $k^2\ll |a''/a|$ corresponding to wavelength
which are outside of the horizon (under the bell)
\begin{equation}
\psi(k,\eta) = A(k) a(\eta) + B(k) a(\eta)\int^{\eta} \frac{d\eta'}{a^2},
\end{equation}
where $A(k)$ and $B(k)$ are integration constants. For large momenta
(i.e.  $k^2\gg |a''/a|$) the general solution of Eq. (\ref{psi}) is
\begin{equation}
\psi(k,\eta) = \frac{1}{\sqrt{2 k}} \bigl[ \beta_{+}(k) e^{- i k \eta} 
+ \beta_{-}(k) e^{i k\eta}\bigr],
\label{plane}
\end{equation}
where $\beta_{-}(k)$ and $\beta_{+}(k)$ are complex 
numbers and where the quantum mechanical normalization
$1/\sqrt{2 k}$ has been chosen.
Therefore, if we  start at $\eta\rightarrow -\infty$ with a 
positive  frequency mode the evolution for $k^2 \ll |a''/a|$ will mix the 
the positive frequency with the negative one 
 and eventually we will end up, for $\eta\rightarrow 
+\infty$ with a superposition of positive and negative 
frequencies according to Eq. (\ref{plane}) 

The two point correlation function of the graviton field operators can then be 
computed in Fourier space
\begin{equation}
\langle 0 | \hat{\mu}_{\lambda}(\vec{k},\eta) 
\hat{\mu}_{\sigma}^{\dagger}(\vec{p},\eta) 
|0\rangle = |\psi_{\lambda}(k,\eta)|^2 \delta_{\lambda\sigma} 
\delta^{(3)}(\vec{k} - \vec{p}).
\label{cor1}
\end{equation}
Field modes with different wave-numbers are then statistically independent 
as a consequence of the graviton emission from the vacuum.
Some authors refer to Eq. (\ref{cor1}) as to the 
{\em stochasticity condition} which express the 
main statistical property of the graviton background.
Within our quantum mechanical formalism we can
also compute the two-point correlation function between 
field operators, namely
\begin{equation}
\xi(r) =\langle 0| \hat{\mu}_{ij}(\vec{x}) 
\hat{\mu}^{ij}(\vec{x} + \vec{r})|0 \rangle=
\frac{1}{\pi^2} \int \frac{d k}{k} \frac{\sin{k r}}{k r}
 |\delta_{h}(k,\eta)|^2,
\label{cor2}
\end{equation}
where 
\begin{equation}
|\delta_{\psi}(k,\eta)|^2 =  k^3 
( |\psi_{\oplus}(k,\eta)|^2 + |\psi_{\otimes}(k,\eta)|^2),
\label{deltapsi}
\end{equation}
is the {\em power spectrum} of the relic gravitons background 
summed over the physical polarizations.

Sometimes, to facilitate the transition to the quantities often used 
by the experimentalists, it is convenient to denote the gravitational 
wave amplitude, for each polarization,  with 
\begin{equation}
h(\vec{x},\eta) = \frac{1}{(2\pi)^{3/2}}\int d^3 k ~h(k,\eta) 
~e^{ i\vec{k}\cdot \vec{x}},
\label{ampl}
\end{equation}
where the reality condition implies $h^{\ast}(\vec{k}) = h(-\vec{k})$.
The statistical independence of waves with different wave vectors 
implies, in this formalism, that
\begin{equation}
\langle h_{\lambda}(\vec{k},\eta) h^{\ast}_{\sigma}(\vec{p},\eta)\rangle = 
|h_{\lambda}(k,\eta)|^2 \delta_{\lambda\sigma} 
\delta^{(3)}(\vec{k} -\vec{p}),~~~h_{\lambda}(k,\eta)
=\frac{\psi_{\lambda}(k,\eta)}{a}.
\label{stoc}
\end{equation}
Eq. (\ref{stoc}) can be viewed as the classical limit of Eq. (\ref{cor1})
where the quantum mechanical expectation values are replaced by ensemble 
averages. Notice that this concept can be made quite rigorous 
by taking the average of the graviton field over the coherent 
state basis. This procedure would be analogous to what is normally done 
in quantum optics in the context of the optical equivalence theorem 
\cite{qo}.

The energy density (and pressure) of the relic graviton background can 
be easily obtained by taking the average of the energy momentum tensor.
The energy density is 
\begin{equation}
\rho_{GW}(\eta) = \langle 0| T_{00}|0\rangle, 
\end{equation}
which can also be written as 
\begin{eqnarray}
&&\rho_{GW}(\eta) = \langle
0_{\otimes}0_{\oplus}|\tau_{oo}|0_{\oplus}0_{\otimes}\rangle=
\nonumber\\
&&\frac{1}{16\pi^3 ~a^2 } \int d^3 k \Biggl[ |h'_{\oplus}(k,\eta)|^2
+
|h'_{\otimes}(k,\eta)|^2 +
k^2 \biggl(|h_{\oplus}(k,\eta)|^2 + |h_{\otimes}(k,\eta)|^2
\biggr)\Biggr].
\label{aven}
\end{eqnarray}
If we insert the  asymptotic expression of the mode function
of each polarization 
for large (positive) conformal times we get, from Eq. (\ref{aven}), 
\begin{equation}
\rho_{GW}(\eta) = \int \frac{d\omega}{\omega}
 \omega^4 \frac{ |\beta_{-}(\omega)|^2 }{\pi^2},
~~~\omega= k/a,
\end{equation}
from which we can define the logarithmic energy spectrum as 
\begin{equation}
\frac{d \rho_{GW}(\omega, \eta)}{d\ln\omega} = 
\frac{\omega^4}{\pi^2} |\beta_{-}(\omega)|^2 ,
\label{endens}
\end{equation}
(where we used the fact that 
$|\beta_{+}(\omega)|^2 - |\beta_{-}(\omega)|^2 =1$)
By taking into account that the energy density of $\overline{n}(\omega)$
gravitons  in the proper momentum interval $d\omega$ is 
given by 
\begin{equation}
d\rho_{GW} = 2 \omega \overline{n}(\omega) \frac{d^{3}\omega}{(2\pi)^3},
\end{equation}
$|\beta_{-}(k)|^2$ can be interpreted as the 
mean number of produced gravitons in a given frequency interval.
 
In order to compute the relation between 
$\Omega_{GW}(\omega,\eta)$ and $\delta_{h}(\omega,\eta)
=\delta_{\psi}(\omega,\eta)/a(\eta)$ we can use Eq. (\ref{ampl}). 
The energy density in gravity waves is given by 
\begin{equation}
\rho_{GW}(\eta) = \frac{M_{P}^2}{6 a^2} \langle {h'}^2_{\otimes} 
+ {h'}^2_{\oplus} \rangle,
\label{enstoc}
\end{equation}
where the brackets indicate now ensemble average of the Fourier amplitudes.
Using now Eq. (\ref{ampl}) together with the stochasticity condition 
we obtain (after having divided by the critical energy density)
\begin{equation}
\Omega_{GW}(\omega, \eta_0) = \frac{1}{6\pi^2} 
\bigl(\frac{\omega}{H_0}\bigr)^2 |\delta_{h}(\omega,\eta_0)|^2.~~~
\label{endensspec}
\end{equation}
where the $\eta_0$ is the present conformal time which appears since we 
want to refer everything to the present value of the Hubble constant.
In comparing the signal coming from a particular model with the 
experimental sensitivities it turns out to be widely used 
the {\em spectral density} $S_{h}(\nu,\eta_0)$. In order to define it 
let us assume that each polarization of the gravity wave amplitude can be 
written as 
\begin{equation}
h(\nu) = \int d t~ h(t)~ e^{-2 i \pi \nu t},
\label{specsens}
\end{equation}
where $t$ is the cosmic time variable.
Then the spectral amplitude can be defined as 
\begin{equation}
\langle h(\nu) h^{\ast}(\nu') \rangle = \frac{1}{2}\delta(\nu-\nu') S_{h}(\nu).
\label{specsens2}
\end{equation}
By inserting this expression into Eq. (\ref{enstoc}) and by using the 
definition (\ref{specsens2}) we obtain 
\begin{equation}
\Omega_{GW}(\nu,\eta_0) = \frac{4\pi^2}{ 3 H_0^2} \nu^3 S_{h}(\nu,\eta_0) .
\end{equation}
This last equation implies that 
\begin{equation}
S_{h}(\nu,\eta_0)= 8\times 10^{-37} \Omega_{GW}(\nu,\eta_0) h_{0}^2 
\frac{ {\rm Hz}^2}{\nu^3}
\end{equation}
from which is clear that $S_{h}$ is measured in seconds. 
\renewcommand{\theequation}{B.\arabic{equation}}
\setcounter{equation}{0}
\section{Relic Gravitons Spectra in the Ordinary Inflationary Case} 

Consider the model of an Universe evolving from a de Sitter stage of 
expansion to a matter-dominated phase passing through a radiation dominated
 phase. By requiring the continuity of the scale factors and of their first 
derivatives at the transition points we have that $a(\eta)$, 
in the three different temporal regions, can be represented as 
\begin{eqnarray}
&& a_{i}(\eta) = \biggl[ -\frac{\eta_1}{\eta}\biggr],~~~{\rm for}
~~~~~~~~~~~~~~~\eta<-\eta_1
\nonumber\\
&& a_{r}(\eta) = \frac{\eta + 2 \eta_1}{\eta_1},~~~{\rm for}
~~ ~~~-\eta_{1} <~ \eta~< 
\eta_{2}
\nonumber\\
&& a_{m}(\eta) = \frac{ (\eta + \eta_{2} + 4\eta_{1})^2}
{4 \eta_{1} ( \eta_2 + 2 \eta_1)},~~~{\rm for}~~~~~~~~\eta >\eta_{2}
\label{ordinf}
\end{eqnarray}
where $-\eta_{1}$ and $\eta_{2}$ mark, respectively, the 
onset of the radiation-dominated phase and the decoupling time.

In order to compute the graviton spectra in this model we have to 
estimate the amplification of the graviton mode function induced by the 
background evolution reported in Eq. (\ref{ordinf}).
The solution of Eq. (\ref{psi}) in the three temporal regions is given by
\begin{eqnarray}
&&\psi_{i}(x) = \frac{p}{\sqrt{2 k}} \sqrt{ x} H_{\nu}^{(2)}( x), 
\nonumber\\
&&\psi_{r}(y) \frac{1}{\sqrt{  2 k}}[ B_{+} e^{- i y} + B_{-} e^{i y}],
\nonumber\\
&&\psi_{m}(z) \frac{1}{\sqrt{2 k}} \sqrt{z} [q^{\ast}  c_{+} H_{\mu}^{(2)}(z)
+ q c_{-} H_{\mu}^{(1)}(z)],
\label{sol}
\end{eqnarray}
where 
\begin{equation}
x= k\eta,~~~ y = k( \eta + 2 \eta_1),~~~z= k(\eta + \eta_2 + 4 \eta_1).
\end{equation}
and $H^{(1,2)}_{\nu, \mu}$ are the Hankel functions \cite{abr,tricomi} of 
first and second kind. 
In the case of the background given by Eq. (\ref{ordinf}) the 
Bessel indeces $\nu$ and $\mu$ are both equal to $3/2$ 
but we like to keep them 
general in light of the applications reported in the present paper.

Notice that in Eq. (\ref{sol}) we included
\begin{equation}
p= \sqrt{\frac{\pi}{2}} e^{-i \frac{\pi}{4}( 1 + 2 \nu)},~~~
q=   \sqrt{\frac{\pi}{2}} e^{i \frac{\pi}{4}( 1 + 2 \mu)},
\end{equation}
ensuring that 
the large time limit of the mode function is the one dictated 
by quantum mechanics (i.e. $e^{i\pm k \eta}/\sqrt{2 k}$) without
any extra phase or extra (constant) coefficient. 

The constants appearing in Eq. (\ref{sol}) are fixed by the quantum
mechanical normalization imposed during the de Sitter phase, and then, 
in order to compute the amplification we have to match the various expressions 
of the mode functions (and of their first derivatives) in $\eta = - \eta_1$ 
and in $\eta=\eta_2$. By doing this we get the expression of the 
amplification coefficients. By firstly looking at modes which went outside 
of the horizon during the de Sitter phase and re-entered during the radiation 
dominated phase (i.e. modes $2\pi/\eta_{1} < k< 2\pi/\eta_{2}$) we have that 
$B_{\pm}$ are given by 
\begin{equation}
B_{\mp} = \frac{e^{ \mp i x_1}}{2} p~\bigl\{H^{(2)}_{\nu}(-x_1)\bigl[
\sqrt{- x_1} \mp \frac{i}{\sqrt{- x_1}} \bigl(\nu + \frac{1}{2}\bigr)\bigr]
\pm i \sqrt{- x_1} H^{(2)}_{\nu + 1}(-x_1)\bigl\}
\end{equation}

We are interested in the $B_{-}$ which measures the amount of mixing between 
positive and negative frequency modes and whose square modulus can be 
interpreted as the mean number of produced gravitons. Moreover, we want to 
evaluate $B_{-}$ in the small argument limit (i.e. $x_1\ll 1$) which is the 
one physically relevant. The result is 
\begin{equation}
B_{-}(k) = \frac{2^{\nu -1}}{\sqrt{ 2 \pi}} \Gamma(\nu)
 \biggl(\frac{1}{2}  - \nu\biggr) e^{-i\frac{\pi}{4} ( 1 + 2 \nu)}
 |k\eta_{1}|^{- \frac{1}{2} - \nu},
\label{bog1}
\end{equation}
where $\Gamma(\nu)$ is the Euler Gamma function.
In the case of a pure de Sitter background we get, indeed, 
\begin{equation}
B_{-}(k) = - \frac{e^{-i\pi}}{2} |k\eta_{1}|^{-2}. 
\end{equation}

In order to compute the spectrum of the relic gravitons 
crossing the horizon during the matter dominated epoch we have to consider 
a further transition by matching $\psi_{r}(y)$ and $\psi_{m}(z)$ (and their 
first derivatives) in $\eta=\eta_2$. The result 
is that 
\begin{equation}
c_{-} = - \frac{i}{2} ( B_{+} - B_{-}) \sqrt{z_2} H^{(2)}_{\mu}(z_2)q,
\label{bog2}
\end{equation} 
where we reported only the dominant part in the $x_1\ll 1$ and $z_2\ll 1$
limit.  By performing the limit explicitly we get
\begin{equation}
|c_{-}(k)|= \frac{\Gamma(\mu)~ \Gamma(\nu)}{2 \pi^2} 
2^{\mu + \nu} \bigl(\frac{1}{2} - \nu\bigr) |q| |p| |k\eta_1|^{- \nu 
-\frac{1}{2}} ~ |k\eta_2|^{ - \mu + \frac{1}{2}}. 
\end{equation}

In the case of the model of Eq. (\ref{ordinf}) we have that the energy 
density of the created gravitons in critical units
at the present observation time 
\begin{equation}
\Omega(\nu, \eta_0) = \frac{1}{\rho_c} \frac{d \rho_{GW}}{d \ln \nu},
\label{def1}
\end{equation}
can be computed by inserting into Eq. (\ref{endens}) the 
expressions of the mixing coefficients of Eqs. (\ref{bog1})--(\ref{bog2}), 
with the result that
\begin{eqnarray}
&&\Omega_{GW}(\nu,\eta_0) \simeq \Omega_{\gamma}(\eta_0) 
\biggl(\frac{H_1}{M_{P}}\biggr)^2,~~~~~~~~~~~\nu_{{\rm dec}} < \nu < \nu_{1} 
\nonumber\\
&& \Omega_{GW}(\nu,\eta_0) \simeq \Omega_{\gamma}(\eta_0) 
\biggl(\frac{H_1}{M_{P}}\biggr)^2 
\biggl(\frac{\nu_{{\rm dec}}}{\nu}\biggr)^2,
~~~\nu_{0}< \nu < \nu_{{\rm dec}} 
\end{eqnarray}
This spectrum is reported in Fig. \ref{f1}. The  {\em soft branch}
and the  {\em hard branch } defined in the Introduction do 
correspond, respectively, to the two frequency ranges 
$\nu_0 <\nu<\nu_{{\rm dec}}$ and $\nu_{{\rm dec}} < \nu <\nu_{1}$. 

\renewcommand{\theequation}{C.\arabic{equation}}
\setcounter{equation}{0}
\section{Transition from Inflation to Stiff Phase: 
Accurate Mixing Coefficients} 

In order to determine the six  matching coefficients appearing in Eq. 
(\ref{solmode}) we have to match the mode function $\psi$ and its 
first derivative in $\eta= -\eta_1$, $\eta = \eta_{r}$ and 
$\eta = \eta_{{\rm dec}}$. The results of this calculation are reported in the 
present section.
Consider first the amplification leading to the {\em hard branch} of the 
spectrum. The mixing coefficients are given by 
\begin{eqnarray}
&&A_{-}(k) = -\frac{\pi}{4\sqrt{2}} e^{ - \frac{i}{2}\pi (\nu + 1) } 
\bigl\{H_{0}^{(2)} \bigl(\frac{x_1}{2}\bigr)\bigl[ x_1 H^{(2)}_{\nu + 1}(-x_1)
- \nu H^{(2)}_{\nu}(-x_1)\bigr] 
- x_1 H^{(2)}_{1}\bigl(\frac{x_1}{2}\bigr) 
H^{(2)}_{\nu}(-x_1) \bigr\}.
\nonumber\\
&&A_{+}(k) = \frac{\pi}{4 \sqrt{2}} e^{ - \frac{i}{2} \pi\nu}\bigl\{ 
H^{(1)}_{0}\bigl(\frac{x_1}{2}\bigr) \bigl[ x_1 H^{(2)}_{\nu + 1}(-x_1)
- \nu H^{(2)}_{\nu}(-x_1)\bigr] 
- x_1 H^{(1)}_{1}\bigl(\frac{x_1}{2}\bigr)
H^{(2)}_{\nu}(- x_1)\bigr\}.
\end{eqnarray}
The small argument limit of $A_{-}$ (directly relevant for the estimate
of the graviton spectrum in the hard branch) can be easily obtained and it 
turns out to be
\begin{equation}
A_{-}(k) \sim \frac{3 \nu}{\pi} 2^{\nu - \frac{3}{2}}e^{- \frac{i}{2}\pi 
( 2 \nu + 1)}\Gamma(\nu) x_{1}^{-\nu} \ln{x_1}.
\label{amin}
\end{equation}
In order to derive the previous and the following expressions it is 
useful to bear  in mind that the Wronskian of the Hankel functions 
is given by
\begin{equation}
{\cal W}\bigl[H^{(1)}_{\alpha}(r), H^{(2)}_{\alpha}(r)\bigr] = 
H^{(1)}_{\alpha+ 1}(r) H^{(2)}_{\alpha}(r) - H^{(1)}_{\alpha}(r) 
H^{(2)}_{\alpha + 1}(r)
= - \frac{4i}{\pi r},
\end{equation}
for a generic argument $r$ and for a generic Bessel index $\alpha$. 
In this calculation it is also useful to recall that the small argument limit 
of the Hankel functions is 
\begin{equation}
H^{(1,2)}_{\alpha}(r) \sim \biggl(\frac{r}{2}\biggr)^{\alpha} 
\frac{1}{\Gamma(\alpha +1)} \mp \frac{i}{\pi} \Gamma(\alpha)  
\biggl(\frac{r}{2}\biggr)^{-\alpha}, 
\end{equation}
where the minus (plus) refers to the Hankel function of first (second) kind.
Notice that this formula is only valid for $\alpha\neq 0$. In this last case 
we have indeed that 
\begin{equation}
H^{(1,2)}_{0}(r)\sim 1 \mp \frac{2}{\pi} \ln{r}.
\end{equation}
Let us now consider the following transition namely the one
leading to the {\em semi-hard branch of the spectrum}. 
The exact expression 
of the mixing coefficients is, in this case,
\begin{eqnarray}
&&B_{\pm}(k) = \frac{e^{ \pm i y_{r}}}{2}\bigl\{ \sqrt{y_r}\bigl[ 
s^{\ast} A_{+}(k)
H^{(2)}_{0}(y_{r}) + s A_{-}(k) H_{0}^{(1)}(y_r)\bigr] \pm i \bigl[
s^{\ast} A_{+}(k) \bigl( \frac{1}{2\sqrt{y_r}} H^{(2)}_{0}(y_r)  -
\sqrt{y_r} H^{(2)}_{1}(y_r)\bigr) 
\nonumber\\
&&+ 
s A_{-}(k) \bigl( \frac{1}{2\sqrt{y_r}} H^{(1)}_{0}(y_r)  -
\sqrt{y_r} H^{(1)}_{1}(y_r)\bigr)\bigr]\bigr\}.
\end{eqnarray}
The small argument limit of $B_{-}$ leads to 
\begin{equation}
B_{-}(k) = - e^{- i \pi \nu + i\frac{\pi}{4}}~ \frac{ 3 \nu }{2 \sqrt{2 \pi}}
2^{\nu - \frac{3}{2}} \Gamma(\nu)~ |k\eta_{r}|^{-\frac{1}{2}} 
~|k\eta_{1}|^{-\nu}\ln{\bigl[\frac{\eta_{1}}{\eta_r}\bigr]},
\label{bmin}
\end{equation}
which becomes, in the  case $\nu = \frac{3}{2}$,
\begin{equation}
B_{-}(k) = \frac{9}{8 \sqrt{2}} ~ |k\eta_{r}|^{-\frac{1}{2}}
~|k\eta_{1}|^{-\frac{3}{2}}\ln{\bigl[\frac{\eta_{1}}{\eta_r}\bigr]}.
\end{equation}

Finally, let us compute the amplification coefficients in the case 
of the {\em soft branch of the spectrum}. 
In this case we have that the mixing coefficients are 
\begin{eqnarray}
&& c_{-}(k) =\frac{i}{2} \bigl\{ \bigl(B_{+}(k) e^{ - iy_{{\rm dec}}} + 
B_{-}(k) e^{  i y_{{\rm dec}}}\bigr) \bigl[ \bigl(\mu + \frac{1}{2}\bigr)
\frac{q^{\ast}}{\sqrt{z_{{\rm dec}}}} H^{(2)}_{\mu}(z_{{\rm dec}}) 
\nonumber\\
&&- \sqrt{z_{{\rm dec}}} q^{\ast} H^{(2)}_{\mu + 1}(z_{{\rm dec}})\bigr]
+ i  \bigl(B_{+}(k) e^{ - iy_{{\rm dec}}} -B_{-}(k) e^{  i y_{{\rm dec}}}\bigr)
\sqrt{z_{{\rm dec}}}q^{\ast} H^{(2)}_{\mu}(z_{{\rm dec}})\bigr\}.
\nonumber\\
&& c_{-}(k) =-\frac{i}{2} \bigl\{ \bigl(B_{+} e^{ - iy_{{\rm dec}}} + 
B_{-}(k) e^{  i y_{{\rm dec}}}\bigr) \bigl[ \bigl(\mu + \frac{1}{2}\bigr)
\frac{q}{\sqrt{z_{{\rm dec}}}} H^{(1)}_{\mu}(z_{{\rm dec}}) 
\nonumber\\
&&- \sqrt{z_{{\rm dec}}}q H^{(1)}_{\mu + 1}(z_{{\rm dec}})\bigr]
+ i  \bigl(B_{+}(k) e^{ - iy_{{\rm dec}}} -B_{-} e^{  i y_{{\rm dec}}}\bigr)
\sqrt{z_{{\rm dec}}}q H^{(1)}_{\mu}(z_{{\rm dec}})\bigr\}.
\end{eqnarray}
In the small argument limit we have 
\begin{equation}
c_{-}(k) = -\frac{3 \nu}{4 \pi} 2^{\mu + \nu-\frac{3}{2}} 
e^{ - i \pi ( \nu + 2 \mu)} \Gamma(\mu)\Gamma(\nu) |k\eta_{r}|^{-\frac{1}{2}}
|k\eta_1|^{-\nu} | k\eta_{{\rm dec}}|^{-\mu +\frac{1}{2}}
 \ln{\bigl[\frac{\eta_{1}}{\eta_r}\bigr]},
\label{cmin}
\end{equation}
which becomes, in the case $\mu= 3/2$,
\begin{equation}
c_{-}(k)= - \frac{9~\pi}{16\sqrt{2}} e^{-i \frac{9}{2} \pi}
|k\eta_1|^{-\frac{3}{2}} | k\eta_{{\rm dec}}|^{-1}|k\eta_{r}|^{-\frac{1}{2}}
 \ln{\bigl[\frac{\eta_{1}}{\eta_r}\bigr]}.
\end{equation}
The derivations  we just reported are the basis for the 
results presented in Section III.

\end{appendix}

\end{document}